\documentclass[fleqn,usenatbib]{mnras}

\usepackage{newtxtext,newtxmath}

\usepackage[T1]{fontenc}

\DeclareRobustCommand{\VAN}[3]{#2}
\let\VANthebibliography\thebibliography
\def\thebibliography{\DeclareRobustCommand{\VAN}[3]{##3}\VANthebibliography}

\usepackage{graphicx}	
\usepackage{amsmath}	

\usepackage{amssymb}	
\usepackage{bbold}      
\usepackage[normalem]{ulem}       
\usepackage{commath}
\usepackage{bm}  
\usepackage{xcolor}

\title[Turbulent Dust Transport in Protoplanetary Disks]{Beyond Diffusion: A Generalized Mean-Field Theory of Turbulent Dust Transport in Protoplanetary Disks}

\author[F. Binkert et al.]{Fabian Binkert,$^{1,2}$\thanks{E-mail: fbinkert@usm.lmu.de}
\\
$^{1}$University Observatory, Faculty of Physics, Ludwig-Maximilians-Universität München, Scheinerstr. 1, 81679 Munich, Germany\\
$^{2}$Exzellenzcluster ORIGINS, Boltzmannstr. 2, D-85748 Garching, Germany}

\date{Accepted XXX. Received YYY; in original form ZZZ}

\pubyear{2022}

\begin{document}
\label{firstpage}
\pagerange{\pageref{firstpage}--\pageref{lastpage}}
\maketitle

\begin{abstract}
Turbulence in protoplanetary disks, when present, plays a critical role in transporting dust particles embedded in the gaseous disk component. When using a field description of dust dynamics, a diffusion approach is traditionally used to model this turbulent dust transport. However, it has been shown that classical turbulent diffusion models are not fully self-consistent. Several shortcomings exist, including the ambiguous nature of the diffused quantity and the nonconservation of angular momentum. Orbital effects are also neglected without an explicit prescription. In response to these inconsistencies, we present a novel Eulerian turbulent dust transport model for isotropic and homogeneous turbulence on the basis of a mean-field theory. Our model is based on density-weighted averaging applied to the pressureless fluid equations and uses appropriate turbulence closures. Our model yields novel dynamic equations for the turbulent dust mass flux and recovers existing turbulent transport models in special limiting cases, thus providing a more general and self-consistent description of turbulent particle transport. Importantly, our model ensures the conservation of global angular and linear momentum unconditionally and implicitly accounts for the effects of orbital dynamics in protoplanetary disks. Furthermore, our model correctly describes the vertical settling-diffusion equilibrium solutions for both small and large particles. Hence, this work presents a generalized Eulerian turbulent dust transport model, establishing a comprehensive framework for more detailed studies of turbulent dust transport in protoplanetary disks. 
\end{abstract}


\section{Introduction}
Protoplanetary disks are believed to exhibit turbulence, driving the redistribution of angular momentum and accretion. Observational constraints on the typical strength of disk turbulence suggest a dimensionless $\alpha$-parameter \citep{Shakura1973} of $10^{-4}$-$10^{-3}$ \citep{Lesur22}. However, the exact nature and origin of disk turbulence remain unclear. Potential sources include (magneto) hydrodynamic instabilities such as the \textit{magnetorotational instability} (MRI) \citep{Balbus91}, which can occur if partly ionized gas in quasi-Keplerian rotation couples to a magnetic field. Purely hydrodynamic instabilities include the \textit{vertical shear instability} (VSI) \citep{Arlt04,Nelson13}, the \textit{convective overstability} \citep{Klahr14, Lyra14} and the \textit{zombie vortex instability} \citep{Barranco05,Lesur16}. The specific operating mechanism depends on the disk structure. \newline
In addition to driving accretion, turbulence also poses an obstacle to the initial stages of planet formation, specifically dust growth and planetesimal formation. In particular, dust grains within protoplanetary disks are aerodynamically coupled to turbulent gas flows, which influences dust growth \citep{Voelk1980,Ormel07,Birnstiel2010}, the dust distribution \citep{Fromang06}, and dust transport \citep{Cuzzi1993,Youdin2007,Carballido10,Zhu15}. Turbulence also prevents dust grains from efficiently clumping together \citep{Umurhan20,Chen20,Gole20} to form planetesimals through mechanisms like the \textit{streaming instability} \citep{Johansen07} because turbulence acts to diffuse particle concentrations \citep{Goodman2000,Youdin05}.\newline
With the Atacama Large Millimeter/submillimeter Array (ALMA) enabling spatially resolved observations of the dust distribution in protoplanetary disks, it has become essential for numerical disk models to incorporate dust physics in addition to gas in order to constrain the physical processes observed in these disks.\newline
While the Navier-Stokes equations effectively describe gas dynamics in such models, no single mathematical tool similarly dominates the description of dust dynamics. Among others, two major approaches to dust modeling in protoplanetary disks are the \textit{Lagrangian} description \citep{Youdin07c,Charnoz11,Yang16,Mignone19}, which describes individual dust particle motion, and the \textit{Eulerian} or \textit{fluid} approach \citep{Johansen05,Paardekooper2006,Meheut12,Benitez-Llambay2018, Huang22}, which describes the collective particle behavior.\newline
Regardless of the approach, solving particle dynamics in protoplanetary disks numerically is especially challenging due to the wide range of spatial and temporal scales involved, particularly in the presence of turbulence. To capture the entire physics of the problem, all relevant length scales must be resolved, which can be computationally demanding and often impossible with current computational capabilities. Further, the detailed nature and origin of turbulence in these disks, if present, is often unknown. Therefore, hydrodynamic dust models are frequently extended with specific phenomenological models that describe the effects of turbulence rather than self-consistently modeling the turbulence itself.\newline
For example, stochastic turbulence models add random fluctuations to the velocities of the gas and dust particles, simulating turbulent mixing and transport. These stochastic models are often used in Lagrangian dust models, allowing accurate modeling of the complex interactions between gas and dust in the presence of turbulence. In contrast, Eulerian turbulence models typically introduce a diffusion term to the dust continuity equation to account for turbulent transport effects \citep[][and Sec.~\ref{sec:The Turbulent Particle Diffusion Model}]{Cuzzi1993,Goodman2000,Dullemond18, Weber20}. However, the classical diffusion approach has a few inconsistencies. Specifically, the approach does not necessarily conserve angular momentum \citep{Tominaga19,Weber20}, and there is no consensus on whether the quantity diffused by turbulence is either the absolute dust density \citep[e.g.][]{Cuzzi1993} or the dust concentration relative to gas \citep[e.g.][]{Dubrulle1995}. Furthermore, the classical diffusion model must be explicitly adapted for applications in Keplerian disks because orbital effects can reduce the strength of diffusivity \citep{Youdin2007}, an effect that is not captured by such diffusion models. 
The aforementioned inconsistencies can be problematic because accurately capturing the physics of turbulent transport is crucial for interpreting observations of protoplanetary disks and their dust distributions, and consequently, for improving our understanding of planet formation.\newline
Recently, two ways have been proposed to resolve the issue concerning the non-conservation of angular momentum. One of which is to introduce correction terms to the dust momentum equation \citep{Tominaga19}, while making sure not to violate Galilean invariance in the process \citep{Huang22}. A second solution was proposed by \citet{Klahr2021}, who modeled turbulent transport with a pressure-like term. \newline
Motivated by the general inconsistencies, we remain agnostic to the specific source of turbulence in this paper, and derive a novel self-consistent Eulerian turbulence model that conserves angular momentum, resolves the question of the fundamental transport quantity, and intrinsically incorporates orbital effects. We recover the previous turbulence models as special limiting cases of our novel turbulent transport model. As such, our approach removes the tension which is currently present in turbulent transport modeling of particles in protoplanetary disks and provides a novel framework for understanding the complex interplay between turbulence and particle dynamics in protoplanetary disks.\newline
The outline of this paper is as follows. We first review relevant theoretical background in Sec.~\ref{sec:Background}, including a brief review of current Eulerian turbulent diffusion models. On the basis of the introduced theory, we derive a novel turbulent transport model in Sec.~\ref{sec:TheTPPM}, and then discuss its applications to dust modeling in turbulent protoplanetary disks in Sec. \ref{sec:Properties and Applications of the TPP Model}. In Sec.~\ref{sec:LPA}, we study the effects of turbulent transport on harmonic perturbations in the absence of external forces and also in the presence of orbital effects. Lastly, in Sec.~\ref{sec:Summary}, we summarize our findings.

\section{Theoretical Background}
\label{sec:Background}
This section first provides an overview of gas and dust dynamics in Secs.~\ref{sec:gas_dynamics} and \ref{sec:Dust Dynamics} respectively. Subsequently, we compare dust modeling via Lagrangian and Eulerian descriptions in Sec.~\ref{sec:gov:equations}. Sec.~\ref{sec:Characteristics of Disk Turbulence} introduces statistical characteristics of turbulence and defines important turbulent transport quantities, such as the \textit{diffusion coefficient} and the \textit{correlation time}. We review a stochastic Lagrangian turbulent dust transport model and the classical Eulerian gradient diffusion model in Sec.~\ref{sec:Stochastic Lagrangian Formalism} and Sec.~\ref{sec:The Turbulent Particle Diffusion Model} respectively. Therein, we also discuss the limitations of applying the latter model to turbulent dust transport in protoplanetary disks. In Sec.~\ref{eq:Mean-Flow Equations}, we introduce the concept of mean-field theory, and briefly review recent work on turbulent dust transport by \cite{Tominaga19}, \cite{Huang22}, and \citet{Klahr2021} in Sec.~\ref{sec:recent_work}. Finally, Sec.~\ref{sec:Hinze-Tchen Model} discusses the turbulent particle dispersion as described by the Hinze-Tchen formalism.

\subsection{Gas Dynamics}
\label{sec:gas_dynamics}
The dynamics of the gaseous component of an inviscid, unmagnetized protoplanetary disk are governed by the time-dependent Euler equations which in Cartesian coordinates read \citep[e.g.][]{Shu92} 
\begin{equation}\label{eq:gas_continuity}
    \frac{\partial \rho_g}{\partial t}+\frac{\partial}{\partial x_j}\big(\rho_g u_{j}\big)=0
\end{equation}

\begin{equation}\label{eq:gas_mom_cons}
        \frac{\partial}{\partial t}(\rho_gu_{i})+\frac{\partial}{\partial x_j}\big(\rho_gu_{i} u_{j}\big)+\frac{\partial}{\partial x_i}p=\rho_g g_i
\end{equation}
where we have used the Einstein summation convention. The \textit{continuity equation} (Eq.~\ref{eq:gas_continuity}) represents the conservation of mass and describes the evolution of the gas volume density $\rho_g$. Here, $u_{i}$ represents is the gas velocity along dimension $i=1,2,3$. Equation \ref{eq:gas_mom_cons} describes the dynamics of the gas momentum per unit volume $\rho_gu_{i}$. While the first and second terms in Eq.~\ref{eq:gas_mom_cons} account for the local change and advection of momentum respectively, the third and fourth terms are contributions by the gradient pressure force and gravitational force respectively. Specifically, $p$ is the gas thermal pressure and $g_i$ is the gravitational acceleration along dimension $i$. In a low-mass protoplanetary disk, the gravitational acceleration is approximately spherically symmetric and points towards the central star with a magnitude g = $G M_* / r^2$, where $G$ is the gravitational constant, $M_*$ is the mass of the star, and $r$ is the distance to the star. \newline
To solve the system of Eqs.~\ref{eq:gas_continuity} and \ref{eq:gas_mom_cons}, an equation of state must be defined. For simple models of protoplanetary disks, a locally isothermal equation of state is often assumed, eliminating the need for an additional energy equation. The locally isothermal equation of state reads:
\begin{equation}\label{eq_isothermal_eos}
    p = \rho_g c_s^2
\end{equation}
where $c_s$ is the isothermal speed of sound and is related to the gas temperature $T$ as
\begin{equation}\label{eq:c_s_T_relation}
    c_s = \sqrt{\frac{k_B T}{m_\mu}}
\end{equation}
where $k_B$ is the Boltzmann constant and $m_\mu$ is the mean mass of a gas molecule. A locally isothermal equation of state assumes a constant gas temperature at all times, which is in many cases a reasonable approximation in protoplanetary disks. \newline
Replacing the gas pressure $p$ in Eq.~\ref{eq:gas_mom_cons} with Eq. \ref{eq_isothermal_eos} closes the system of equations.\newline
Next, we consider the vertical static equilibrium solution to the Euler equations in protoplanetary disks. Static equilibrium solutions require the time derivatives and velocities $u_i$ to vanish, which trivially satisfies the continuity equation (Eq.~\ref{eq:gas_continuity}). In a vertical static equilibrium, the gradient pressure force and the gravitational force must balance exactly in the vertical direction (along the z-axis):
\begin{equation}
        \frac{\partial p}{\partial z}=\rho_g g_z
\end{equation}
We restrict the analysis to regions close to the disk midplane where the $z$-component of the stellar gravitational field can be approximated, based on geometrical arguments, as \citep[e.g.][]{Armitage2009}
\begin{equation}\label{eq:vertical_grac_acc}
    g_z = -\Omega_K^2z
\end{equation}
where $z$ is the distance to the disk midplane and $\Omega_K$ is the \textit{Keplerian angular velocity}  $\Omega_K = \sqrt{G M_* / r^3}$. \newline
Assuming the sound speed is vertically constant, Eq.~\ref{eq:gas_mom_cons} simplifies to describe the vertical disk structure as: 
\begin{equation}\label{eq:vertical_structure}
    c_s^2 \frac{\partial\rho_g}{\partial z} = -\Omega_K^2z
\end{equation}
Integrating Eq.~\ref{eq:vertical_structure} gives the gas volume density as a function of the distance to the disk midplane
\begin{equation}\label{eq:vertical_gas_profile}
    \rho_g=\rho_{g,0}\exp\bigg(-\frac{z^2}{2h_g^2}\bigg)
\end{equation}
where $\rho_{g,0}$ is a constant, and we have defined the vertical \textit{gas pressure scale height} $h_g$ as the ratio between the sound speed and the Keplerian angular velocity 
\begin{equation}\label{eq:sch_defintion}
    h_g\equiv\frac{c_s}{\Omega_K}.
\end{equation}
The \textit{surface density} $\Sigma_g$ is the integral of the volume density along the $z$-axis
\begin{equation}
   \Sigma_g=\int^{+\infty}_{-\infty}\rho_g\textrm{d}z
\end{equation}
and is related to the constant $\rho_{g,0}$ in Eq.~\ref{eq:vertical_gas_profile} as
\begin{equation}
   \rho_{g,0} = \frac{\Sigma_g}{\sqrt{2\pi} h_g}
\end{equation}

\subsection{Dust Dynamics}
\label{sec:Dust Dynamics}
Dust dynamics in protoplanetary disks play an important role in the initial stages of planet formation. It is the micron-sized grains, inherited from the interstellar medium (ISM) from which all rocky bodies in planetary systems grow (e.g., planetesimals, terrestrial planets, cores of giant planets). Additionally, these dust grains significantly contribute to the opacity, and they facilitate chemical surface reactions that can lead to the formation of complex molecules.  \newline
Unlike gas dynamics which are influenced by pressure forces, dust dynamics are predominantly dictated by aerodynamic drag, thereby coupling dust particles to the motion of the gas. The degree of coupling is characterized by the \textit{stopping time} $t_s$, which is the characteristic time in which relative velocities between dust and gas decay due to aerodynamic drag. Assuming spherical dust grains of size $a$ and constant solid density $\rho_\bullet$, the stopping time can be expressed as \citep{Whipple1972,Weidenschilling1977}
\begin{equation}\label{eq:st_time}
    t_s = \sqrt{\frac{\pi}{8}}\frac{a \rho_\bullet}{c_s \rho_g}
\end{equation}
which indicates that small dust grains are more strongly coupled than large grains and that coupling is stronger in a high-gas-density environment compared to a low-density environment. It should be noted that Eq.~\ref{eq:st_time} is valid only for dust grains smaller than the mean free path between individual gas molecules, typical in protoplanetary disk environments.\newline
For applications in protoplanetary disks, the stopping time is typically normalized by the Keplerian angular velocity $\Omega_K$ which then gives the dimensionless \textit{Stokes number}
\begin{equation}\label{eq:Stokes_N_definition}
    St = t_s \Omega_K 
\end{equation}
The force per unit volume exerted by aerodynamic drag on dust is proportional to the relative velocity between dust and gas ($v_{i}-u_{i}$): 
\begin{equation}\label{eq:drag_force_density}
    f_i^\mathrm{drag}=-\frac{\rho_d}{t_s}(v_i-u_i)
\end{equation}
This generates an equal but opposite \textit{back reaction} force on the gas. In most protoplanetary disk environments, the back reaction is negligible, given the local dust density $\rho_d$ is typically much smaller than the local gas density $\rho_g$. However, specific transport mechanisms can cause the dust density to locally increase relative to gas density, making the back reaction relevant again. \newline
An example of such a mechanism is vertical settling. Dust particles suspended above the disk midplane experience a vertical gravitational force (Eq.~\ref{eq:vertical_grac_acc}) accelerating them towards the disk midplane. Strong aerodynamic coupling quickly counterbalances this gravitational force with counteracting drag. By equating gravity (Eq.~\ref{eq:vertical_grac_acc}) and drag (Eq.~\ref{eq:drag_force_density}) forces and rearranging terms, we derive the vertical dust settling velocity \citep{Dubrulle1995}:
\begin{equation}\label{eq:vertical_sett_velocity}
    v_\mathrm{sett}=-t_s\Omega_K^2 z
\end{equation}
which represents the characteristic velocity of dust grains settling towards the disk midplane and depends on the dust grain size via the stopping time $t_s$. It is important to note that Eq.~\ref{eq:vertical_sett_velocity} is valid only when the time taken by a dust particle to reach terminal velocity is negligible compared to the settling time, a condition known as the \textit{terminal velocity approximation}. The terminal velocity approximation strictly holds only for small particles ($St\ll 1$).

\subsection{Lagrangian and Eulerian Dust Modeling}
\label{sec:gov:equations}
This section presents two prevalent mathematical descriptions for modeling dust dynamics in protoplanetary disks, namely the Lagrangian and Eulerian descriptions.\newline
Lagrangian models trace the motion of individual dust particles as they interact with the gas and potentially other dust particles in the disk. The particle trajectories are described by ordinary differential equations, which fully capture the discrete nature of dust particles. Assuming a purely deterministic trajectory, the equations of motion are most effectively described by a Newtonian formalism:
\begin{equation}\label{eq:lagr_1}
    \frac{d x_i}{dt} = v_{i}
\end{equation}
\begin{equation}\label{eq:lagr_2}
    \frac{d v_{i}}{dt} = -\frac{1}{t_s}\big(v_{i}-u_{i}\big) + g_i
\end{equation}
The above equations describe the rate of change of the particle position $x_i$ and velocity $v_{i}$. The r.h.s. of Eq.~\ref{eq:lagr_2} contains the drag force and the gravitational acceleration $g_i$. Because the drag force term contains the gas velocity $u_{i}$, gas dynamics must be known and solved concurrently with particle dynamics.\newline
A downside to the Lagrangian approach is its computational cost, which scales with the number of particles in the model. Typically, numerical models include a significantly smaller number of particles than the physical particle count in protoplanetary disks, representing these particles as \textit{super particles} \citep[e.g.][]{Youdin2007,Zsom08,Wafflard20}. An additional challenge is balancing the large computational demands of regions with high particle density against the limited resolution in low-density areas.\newline
Conversely, when individual particle trajectories are irrelevant to a specific problem, and a field description of fundamental flow properties is appropriate (e.g., in terms of mass flux and concentrations), the Eulerian continuum approach can be a suitable alternative. This method constructs \textit{fluid elements} containing a sufficient number of particles to allow volume-averaged quantities like temperature, density, and velocity to statistically describe each fluid element. Nonetheless, the fluid elements must be small relative to the characteristic lengths scale of the system. Usually, a \textit{grid} best represents these fluid elements, subdividing the domain of interest into individual \textit{cells}.\newline
Assuming an appropriate grid exists, the dust phase of a protoplanetary disk can be described by continuum equations analogous to the gas's Navier-Stokes equations. Typically, a set of Euler-like equations in the limit of vanishing particle dispersion describes a pressureless fluid, an appropriate approximation for particles well-coupled to the gas ($St\ll 1$).\newline
For brevity, we will discuss a single-sized particle population, although the approach can be readily generalized \citep[see e.g.][]{Benitez-Llambay2018}. In conservation form, the pressureless fluid equations read:
\begin{equation} \label{eq:mass_cons}
    \frac{\partial \rho_d}{\partial t}+\frac{\partial}{\partial x_j}\big(\rho_d v_{j}\big)=0
\end{equation}
\begin{equation}\label{eq:mom_cons}
        \frac{\partial}{\partial t}(\rho_dv_{i})+\frac{\partial}{\partial x_j}\big(\rho_dv_{i} v_{j}\big)=-\frac{\rho_d}{t_s}\big(v_{i}-u_{i}\big)+\rho_d g_i
\end{equation}
The pressureless equations are derived from the conservation of mass and momentum, respectively \cite[e.g.][]{Fan1998}. The r.h.s. of Eq.~\ref{eq:mom_cons} models the momentum exchange through aerodynamic interactions of the particles with the gas and gravity. Because the particle dispersion vanishes, there is no need for an additional particle energy equation. \newline
In both the Lagrangian and the Eulerian description, particles couple to gas motion via the drag term. If the gas flow is turbulent, the particles couple to the turbulent flow through this term, making additional turbulence models for dust redundant (assuming the turbulent flow in gas is fully characterized). However, the nature of turbulence in protoplanetary disks often remains unknown or requires very large temporal and spatial resolution to fully capture \citep[e.g.][]{Manger20}. The following sections will discuss the profound impact of turbulence on both gas and dust dynamics.

\subsection{Statistical Characteristics of Turbulence}
\label{sec:Characteristics of Disk Turbulence}
There exists no universal turbulence model for protoplanetary disks. Therefore, turbulent fluctuations are typically characterized statistically and compared against specific models, experiments, and observations. \newline
In our statistical analysis, we follow \citet{Fan1998} and use a Lagrangian tracer that follows the turbulent dynamics of a gas fluid parcel. We simplify by assuming isotropic and homogeneous turbulence, which allows us to describe the turbulent displacement of the fluid parcel along a single dimension. Given the initial position of the fluid parcel $x=0$ at time $t=0$ and assuming the turbulent velocity fluctuation $u'$ is known at all times, the position at times $t>0$ can be evaluated as:
\begin{equation}
    x(t)=\int_0^tu'(t')\mathrm{d}t'
\end{equation}
The averaged squared displacement $\overline{x^2}$ of a fluid parcel subject to statistically steady turbulence, is related to the autocorrelation function of the turbulent velocity fluctuations as \citep{Taylor20}
\begin{equation}\label{some_equation_with_Int}
    \overline{x^2}(t)= 2\int_0^t\mathrm{d}t' \int_0^{t'} \mathrm{d}\tau\:\overline{u'(\tau)u'(0)}
\end{equation}
Here, the overbar signifies a statistical ensemble average. We then define the diffusion coefficient $D$ as the averaged growth rate of the squared displacement over long times \citep[e.g.][]{Fan1998}:
\begin{equation}\label{def:diffusion_coefficient}
    D \equiv \frac{1}{2}\frac{\mathrm{d}\overline{x^2}}{\mathrm{d}t}
\end{equation}
In a purely diffusive process, Eq.~\ref{def:diffusion_coefficient} approaches a constant value.\newline 
According to the \textit{Wiener-Kinchin theorem}, the energy spectrum of turbulent fluctuations in frequency space $\hat{E}_g(\omega)$ is related to the autocorrelation function of the turbulent velocity fluctuations via its Fourier transform:
\begin{equation}\label{eq:Energy_spectrum_def}
    \hat{E}_g(\omega)=\frac{1}{2\pi}\int_{-\infty}^\infty \mathrm{d}t' \: \overline{u'(t')u'(0)}e^{i\omega t'}
\end{equation}
The above equation suggests that in statistically steady turbulence, the energy spectrum is an even function of frequency $\hat{E}_g(\omega)=\hat{E}_g(-\omega)$ \citep[e.g.][]{Zhu15}. The gas diffusion coefficient $D_g$ is then expressed as the integral over the autocorrelation function \citep[e.g.][]{Youdin2007}:
\begin{subequations}\label{eq:base_equation}
\begin{alignat}{4}
    D_g &= \int_0^\infty \mathrm{d}t'\:\overline{u'(t')u'(0)} \label{eq:D_a}\\
    &= \int_0^\infty \mathrm{d}t' \int_{-\infty}^\infty d\omega \:\hat{E}_g(\omega)e^{-i\omega t'}\label{eq:D_b}\\
    & = \pi \int_{-\infty}^\infty \mathrm{d}\omega \: \hat{E}_g(\omega) \delta(\omega)\label{eq:D_c}\\
    &=\pi \hat{E}_g(0)\label{eq:D_d}
\end{alignat}
\end{subequations}
The second line (Eq.~\ref{eq:D_b}) follows from Eq.~\ref{eq:Energy_spectrum_def}. The third line, i.e., Eq.~\ref{eq:D_c}, introduces the delta distribution as the Fourier transformation of a constant and makes use of the fact that the energy spectrum $\hat{E}_g(\omega)$ is an even function to extend the lower integration boundary to $-\infty$. This shows that the diffusion coefficient $D_g$ is proportional to the energy spectrum at $\omega=0$.\newline
We introduce the \textit{correlation time} of turbulence $t_\mathrm{corr}$:
\begin{equation}\label{eq:def_of_t_corr}
    t_\mathrm{corr} \equiv\int_0^\infty \mathrm{d}t' \: \frac{\overline{{u'}(t'){u'}(0)}}{\overline{{u'}^2}}
\end{equation}
Here, $\overline{{u'}^2}=\overline{u'(0)u'(0)}$ is the square of the mean turbulent velocity dispersion. \newline
With the correlation time $t_\mathrm{corr}$ and the diffusion coefficient $D_g$ as the two important statistical characteristics of turbulence, we use Eq.~\ref{eq:D_a} to connect these two quantities with the turbulent velocity dispersion:
\begin{equation}\label{eq:charact_relation_D_t_corr} 
    D_g = \overline{{u'}^2}\:t_\mathrm{corr}
\end{equation}
These definitions apply for any form of the energy spectrum $\hat{E}(\omega)$, provided the turbulence is statistically steady, homogeneous, and isotropic.\newline
We also define a characteristic \textit{eddy length} of the turbulence
\begin{equation}\label{eq:eddy_length}
    l_\mathrm{eddy}\equiv \sqrt{\:\overline{{u'}^2}}\:t_\mathrm{corr}
\end{equation}
and the \textit{diffusion timescale}
\begin{equation}\label{eq:diffusion timescale}
    t_\mathrm{diff}\sim \frac{l^2}{D_g}
\end{equation}
which describes the time for a fluid parcel to diffuse across a distance~$l$.\newline
In the literature, it is common to relate the gas diffusion coefficient $D_g$ to the particle diffusion coefficient $D_d$ via the dimensionless \textit{Schmidt number} \citep[e.g.][]{Cuzzi1993}:
\begin{equation}\label{eq:def_Schmidt_number}
    Sc \equiv \frac{D_g}{D_d}
\end{equation}
In the absence of external forces, $Sc=1$ holds, making the diffusion of dust indistinguishable from that of gas \citep{Youdin2007}. Hereafter, we will use $D$ to represent both $D_g$ and $D_d$ when distinction is not required.\newline
Some studies define the Schmidt number differently, as $Sc_\mathrm{hydro}=\nu/D_g$, the ratio between the kinematic viscosity $\nu$, and $D_g$ \citep[e.g.][]{Johansen05,Carballido06}. Thus, $Sc_\mathrm{hydro}$ quantifies the relative effectiveness of angular momentum transport (associated with $\nu$) and mixing processes (associated with $D_g$). The two parameters $\nu$ and $D_g$ have the same dimensions and both arise from the same turbulence and are therefore closely related but not necessarily equal \citep[see e.g.][]{Pavlyuchenkov07}. Consequently, $Sc_\mathrm{hydro}$ represents a related, albeit not necessarily equivalent, quantity to our definition. In this paper, we adopt the definition of the Schmidt number as given in Eq.~\ref{eq:def_Schmidt_number}, consistent with the convention in \cite{Cuzzi1993} and \citet{Youdin2007}.\newline
When studying protoplanetary disks, turbulent diffusion is sometimes parametrized using a dimensionless \textit{diffusivity} parameter $\delta$:
\begin{equation}\label{eq:turbulent_delta}
    \delta = \frac{D}{c_s h_g}
\end{equation}
The definition in Eq.~\ref{eq:turbulent_delta} is analogous to the dimensionless $\alpha$-parameter (see Eq.~\ref{eq:alpha_parametrization}) introduced by \cite{Shakura1973}. However, while $\delta$ parametrizes the level of turbulent diffusion, the $\alpha$ parameter is most commonly used to describe the efficiency of angular momentum transport in a disk. For $Sc_\mathrm{hydro}=1$, we find $\delta = \alpha$. \newline

\subsection{Stochastic Lagrangian Formalism}
\label{sec:Stochastic Lagrangian Formalism}
In the context of turbulent dust dynamics, a Lagrangian description can incorporate a stochastic forcing term into Eq.~\ref{eq:lagr_2}, such that the turbulent velocity fluctuation readily fulfills the desired turbulence statistics as discussed in Sec.~\ref{sec:Characteristics of Disk Turbulence}.\newline
When applying a stochastic Lagrangian turbulence model, the gas velocity $u$ is typically decomposed into a mean-field contribution $\bar{u}$ and a turbulent fluctuation $\delta u$ such that $u=\bar{u}+\delta u$. The Lagrangian velocity equation for the dust, in one dimension, then becomes:
\begin{equation}\label{eq:lagr_3}
    \frac{d v}{dt} = -\frac{1}{t_s}\big(v-\bar{u}\big) + g+\frac{1}{t_s}\delta u
\end{equation}
The r.h.s. of Eq.~\ref{eq:lagr_3} includes terms for drag and gravity, with $\bar{u}$ signifying the laminar gas velocity \citep[e.g.][]{Charnoz11}. The last term represents the acceleration of dust via the turbulent gas velocity field $\delta u$.\newline
As mentioned in Sec.~\ref{sec:Characteristics of Disk Turbulence}, isotropic and homogeneous turbulence can be statistically characterized using two parameters, namely the diffusion coefficient $D$ and the correlation time $t_\mathrm{corr}$. Consequently, a desirable stochastic turbulence model should also be parametrized by these two quantities. We adopt the model from \cite{Ormel2018} that explicitly defines the stochastic forcing term. Below, we review their formalism in one dimension, but it can readily be extended to higher dimensions. In addition to Eq.~\ref{eq:lagr_1} and Eq.~\ref{eq:lagr_3}, the model incorporates an equation for the turbulent velocity field and a stochastic differential equation:
\begin{subequations}
\begin{equation}\label{eq:SEOM2}
    \delta u = \sqrt{\frac{D}{t_\mathrm{corr}}}\zeta_t
\end{equation}
\begin{equation}\label{eq:SEOM3}
    \mathrm{d}\zeta_t = -\frac{\zeta_t}{t_\mathrm{corr}}\mathrm{d}t+\sqrt{\frac{2}{t_\mathrm{corr}}}\mathrm{d}W_t
\end{equation}
\end{subequations}
Equation~\ref{eq:SEOM2} contains the turbulent velocity dispersion, parametrized by the diffusion coefficient and the correlation time $\sqrt{\delta u ^2} =\sqrt{D/t_\mathrm{corr}}$, as suggested by Eq.~\ref{eq:charact_relation_D_t_corr}, and a dimensionless stochastic variable $\zeta_t$. The dynamics of the stochastic variable $\zeta_t$ is governed by Eq.~\ref{eq:SEOM3} which formally describes an Ornstein–Uhlenbeck process \citep[][]{Uhlenbeck1930}, where $W_t$ denotes the Wiener process. The differential of the Wiener process is $\mathrm{d}W_t=\sqrt{\mathrm{d}t}\mathcal{N}(0,1)$, where $\mathcal{N}(0,1)$ is the normal distribution with zero mean and unit variance. Hence, $\overline{\zeta_t}=0$ and $\overline{\zeta_t^2}=1$ hold. While $\zeta_t$-values are correlated for timescales shorter than $t_\mathrm{corr}$, they become uncorrelated and normally distributed for longer timescales.\newline
In essence, this model incorporates turbulence as an additional stochastic forcing term in the velocity equation. Being a specific turbulence model, it may not necessarily correspond to actual turbulent processes in protoplanetary disks. However, it has proven immensely useful, because the model is parametrized by only two parameters ($D$, $t_\mathrm{corr}$) and has the desired statistical characteristics of turbulence as discussed in Sec.~\ref{sec:Characteristics of Disk Turbulence}.\newline
In the \textit{strong coupling approximation}, where the stopping time~$t_s$ is small, \citet{Ormel2018} show that the system of equations can be represented by a single stochastic differential equation:
\begin{equation}
    \mathrm{d}x = v \mathrm{d}t+\sqrt{2 D} \mathrm{d}W_t
\end{equation}
This equation is frequently used to model Lagrangian dust transport in turbulent protoplanetary disks \citep[e.g.][]{Ciesla2010,Zsom11,Charnoz11,Krijt16b}. \newline
It is crucial to note that the model as presented in this paper is strictly applicable only in an unstratified gas background. For variations in gas density, additional corrections are necessary, as detailed in \citet[][]{Ormel2018}.\newline
Overall, the stochastic model of \citet{Ormel2018} provides a versatile Lagrangian turbulence model, enabling the simulation of dust dynamics in turbulent environments. In the following sections, we will turn to the Eulerian description.

\subsection{Gradient Diffusion and its Limitations}
\label{sec:The Turbulent Particle Diffusion Model}
In this section, we review the gradient diffusion model, as an example of an Eulerian turbulent transport model. Gradient diffusion is probably the most popular model employed to describe turbulent dust transport in protoplanetary disks. \newline
We will start by introducing the concept of a gradient diffusion flux in Sec.~\ref{sec:Introducing a Gradient Diffusion Flux} and then proceed to highlight several limitations inherent to the gradient diffusion model. Specifically, in Sec.~\ref{sec:The Diffused Quantity}, we will highlight that there appears to be no clear consensus on the functional form of the diffused quantity. 
In Sec.~\ref{sec:Vertical Settling-Diffusion Equilibrium} we will discuss the predictions and limitations of the gradient diffusion model regarding the vertical settling-diffusion equilibrium solution in protoplanetary disks, before we will illustrate issues regarding momentum conservation in Sec.~\ref{sec:Momentum Conservation}. Lastly, in Sec.~\ref{sec:Orbital_Effects}, we will review how the model must be explicitly adapted to incorporate the effects of orbital dynamics in disks. 

\subsubsection{Introducing a Gradient Diffusion Flux}
\label{sec:Introducing a Gradient Diffusion Flux}
When relying on an Eulerian description, i.e., describing the dust component in a turbulent protoplanetary disk as a continuous fluid, Reynolds averaging techniques are typically employed to incorporate turbulent transport effects into the equations describing dust dynamics \citep{Champney90,Cuzzi1993}. This approach introduces an additional turbulent transport flux $J_i$ to the dust continuity equation
\begin{equation} \label{eq:trad_adviff_eq}
    \frac{\partial \rho_d}{\partial t}+\frac{\partial}{\partial x_j}\big(\rho_dv_{j}\big)=-\frac{\partial}{\partial x_j}\big(J_j\big)
\end{equation}
but does not simultaneously predict its functional form. \newline
As discussed in Sec.~\ref{sec:Characteristics of Disk Turbulence}, random turbulent displacements in homogeneous and isotropic turbulence behave diffusively. Unsurprisingly, diffusion approaches have been successful in describing the functional form of the turbulent mass flux $J_i$. The most common approach involves the \textit{gradient diffusion hypothesis}, which assumes the turbulent mass flux $J_i$ to be proportional to the gradient of the particle density, 
\citep[e.g.][]{Cuzzi1993,Goodman2000,Schrapler04,Shariff11}
\begin{equation} \label{eq:trad_adviff_eq_2}
    J_i=-D\frac{\partial}{\partial x_i}\rho_d
\end{equation}
With this functional form, the continuity equation takes the form of an \textit{advection-diffusion equation}. \newline
In applications involving a nonuniform gaseous background density, the turbulent mass flux from Eq.~\ref{eq:trad_adviff_eq_2} is often modified to account for gradients in the gas density:
\begin{equation} \label{eq:trad_adviff_eq_3}
    J_i=-D\rho_g\frac{\partial}{\partial x_i}\frac{\rho_d}{\rho_g}
\end{equation}
This alteration is usually motivated by the heuristic \textit{good mixing condition}, which states that both particle and gas distributions evolve towards a common maximum entropy distribution in which the spatial gradient of the particle concentration vanishes $\rho_d/\rho_g=\mathrm{constant}$ \citep[see e.g.][for a more detailed discussion]{Charnoz11}. \newline
Consequently, the following advection-diffusion equation is predominantly employed to model the turbulent mass transport of particles in protoplanetary disks \citep[e.g.][]{Dubrulle1995,Takeuchi02,DullemondD04,Schrapler04,Fromang06,Ciesla09,Dullemond18}.
\begin{equation} \label{eq:turb_diff_model}
    \frac{\partial \rho_d}{\partial t}+\frac{\partial}{\partial x_j}\big(\rho_dv_{j}\big)=\frac{\partial}{\partial x_j}\bigg[D\rho_g\frac{\partial}{\partial x_i}\bigg(\frac{\rho_d}{\rho_g}\bigg)\bigg]
\end{equation}
Despite the success of the gradient diffusion model in modeling turbulent particle transport in protoplanetary disks, it has certain inherent limitations, which we will discuss in the following sections.

\subsubsection{The Diffused Quantity}
\label{sec:The Diffused Quantity}
The protoplanetary disk community seems not to have reached a consensus on the functional form of the quantity diffused by turbulence. More rigorous mathematical derivations typically result in a diffusion flux, in which the absolute particle density $\rho_d$ is the diffused quantity \citep[e.g.][]{Cuzzi1993,Laibe20}. However, this appears to be inconsistent with the good mixing condition in the small particle limit.\newline
In contrast, heuristic arguments favor the functional form expressed in Eq.~\ref{eq:trad_adviff_eq_3} which assumes the \textit{dust-to-gas ratio} $\rho_d/\rho_g$ to be the diffused quantity, which additionally accounts for gradients in the gas density \citep[e.g.][]{Dubrulle1995,Charnoz11}. Although the latter quantity appears to be the more favorable choice, a self-consistent model supporting this choice is yet to be definitively established. To date, we only know of \cite{Riols18} who have proposed a mathematically coherent argument on the basis of Reynolds averages and assuming strongly coupled particles and small dust concentrations. 

\subsubsection{Vertical Settling-Diffusion Equilibrium}
\label{sec:Vertical Settling-Diffusion Equilibrium}
We now explore the use of the gradient diffusion model in the form of Eq.~\ref{eq:turb_diff_model} to describe the vertical steady-state structure of a protoplanetary disk. For this, we assume a vertically isothermal gaseous background with a vertical hydrostatic equilibrium profile as given by Eq.~\ref{eq:vertical_gas_profile}. Additionally, we assume the background gas to exhibit isotropic, homogeneous turbulence and a constant diffusion coefficient $D$. In a steady state, the particle component in this background is in a vertical settling-diffusion equilibrium, typically found by using the terminal velocity approximation (Eq.~\ref{eq:vertical_sett_velocity}).\newline
Substituting Eq.~\ref{eq:vertical_sett_velocity} into Eq.~\ref{eq:turb_diff_model}, and assuming a steady state such that the time derivative vanishes, we find the following one-dimensional differential equation: 
\begin{equation}\label{eq:diff_z}
    \frac{\partial}{\partial z}\bigg(\ln \frac{\rho_d}{\rho_g} \bigg)=-\frac{\Omega^2 t_s}{D}z
\end{equation}
Further assuming the vertical gas density profile follows the Gaussian profile of Eq.~\ref{eq:vertical_gas_profile} with scale height $h_g$, integration of Eq.~\ref{eq:diff_z} yields \citep{Fromang2009}:
\begin{equation}\label{eq:FromangandNelson}
    \rho_d=\rho_{d,0}\exp\bigg[ -\frac{\Omega t_{s,\mathrm{mid}}}{\delta} \bigg( \exp\bigg( \frac{z^2}{2h_g^2}\bigg)-1\bigg)-\frac{z^2}{2h_g^2}\bigg]
\end{equation}
Here, $t_{s,\mathrm{mid}}$ denotes the stopping time evaluated at the disk midplane, and we have used Eq.~\ref{eq:turbulent_delta} to simplify the expression.  \newline
The above solution relies on the terminal velocity approximation, which neglects inertial accelerations, and thus is only applicable for $St \ll 1$, a regime where drag forces are dominant \citep{Youdin05}. {For small particles, with $St\ll1$ at the disk midplane, this condition is fulfilled everywhere in the disk except the disk atmosphere where even the Stokes number of the smallest particles exceeds unity ($St\gtrsim 1$) due to the exponential stratification of the gas background.} For large particles ($St > 1$), this condition is not fulfilled anywhere and Eq.~\ref{eq:FromangandNelson} technically speaking not applicable. \newline
As \cite{Laibe20} noted, an analytical model predicting the transition from drag-dominant to gravity-dominant dynamics in turbulent protoplanetary disks does not currently exist.

\subsubsection{Momentum Conservation}
\label{sec:Momentum Conservation}
Incorporating a diffusion flux in the continuity equation as described by Eq.~\ref{eq:turb_diff_model} may violate the conservation of linear and angular momentum \citep{Goodman2000,Weber20,Tominaga19}. This non-conservation can be problematic, especially since accurate accounting of angular momentum is key for mass transport in accretion disks. Moreover, \citet{Tominaga19} showed that the non-conservation non-physically changes the properties of the \textit{secular gravitational instability} \citep[e.g.][]{Youdin11}.\newline
Here, we follow \cite{Weber20} to illustrate the non-conservation of linear momentum by combining the velocity equation of a particle fluid
\begin{equation}\label{eq:vel_eq}
        \frac{\partial v_{i}}{\partial t}+\frac{\partial}{\partial x_j}(v_{i} v_{j})=\frac{1}{t_s}(u_{i}-v_{i})
\end{equation}
and the continuity equation including the turbulent mass flux (Eq.~\ref{eq:trad_adviff_eq}) to write the particle momentum equation in conservation form:
\begin{equation}\label{eq:mom_cons_form_1}
        \frac{\partial}{\partial t}(\rho_dv_{i})+\frac{\partial}{\partial x_j}(\rho_dv_{i} v_{j})=\frac{\rho_d}{t_s}(u_{i}-v_{i})-v_{i}\frac{\partial}{\partial x_j}J_j
\end{equation}
The first term on the r.h.s models the acceleration due to aerodynamic drag and exchanges momentum between the gas and particle fluid. The second term, associated with turbulent mass transport, generally cannot be expressed as a divergence term. As such, it can contribute to the non-conservation of dust momentum $\rho_d v_{i}$. \newline
{We want to stress here that the non-conservation of the dust momentum in Eq.~\ref{eq:mom_cons_form_1} by itself is not necessarily a problem. Often, the effect of the last term on the r.h.s. of Eq.~\ref{eq:mom_cons_form_1} is regarded as coming from the turbulent gas-particle interaction. The problem only arises if one considers the full system of dust \textit{and} gas. In the full system, momentum is expected to be conserved, but, there is generally no term analogous to the last term in Eq.~\ref{eq:mom_cons_form_1}, in the gas momentum equations that would model the back reaction of the turbulent gas-particle interaction. Consequently, momentum in the full system is not necessarily conserved.}\newline
To prevent this issue, \cite{Goodman2000} introduce an artificial term to the particle momentum equation. For demonstrative purposes, we follow their approach here and add a term of the form 
\begin{equation}\label{eq:artifical_term}
    ...-J_j\frac{\partial v_{i}}{\partial x_j}
\end{equation}
to the r.h.s. of Eq.~\ref{eq:mom_cons_form_1} such that the momentum equation can be rewritten as
\begin{equation}\label{eq:mom_cons_form_2}
        \frac{\partial}{\partial t}(\rho_dv_{i})+\frac{\partial}{\partial x_j}\big(\rho_dv_{i}v_{j}+v_{i}J_j\big)=\frac{\rho_d}{t_s}(u_{i}-v_{i})
\end{equation}
Now, the diffusion flux is included in the divergence term on the l.h.s., which means the particle momentum is globally conserved in the dust fluid even in the presence of a turbulent mass flux. However, as mentioned before, while it is expected that momentum is conserved in the whole system, there is no clear reason why momentum should be conserved within the dust fluid and cannot be exchanged with the gas. The naive addition of this artificial term (Eq.~\ref{eq:artifical_term}) is unproblematic only if the term equals zero. This condition is met if the velocity gradient in the direction of the turbulent mass transport, or equivalently the dot product between the turbulent mass flux $J_i$ and the gradient of the velocity $\partial v_{i}/\partial x_j$, vanishes.\newline 
Conversely, the approach of adding a diffusion flux to the continuity equation, as in Eq.~\ref{eq:turb_diff_model}, can violate momentum conservation if diffusive transport occurs in the direction of a non-zero velocity gradient, specifically wherever $\partial v_{i}/\partial x_j\neq 0$. This is in agreement with the analysis of \cite{Tominaga19}, who show the non-conservation of angular momentum for diffusive particle transport in the radial direction of a protoplanetary disk, where a radial velocity gradient exists due to the Keplerian shear $\partial v_{\phi}/\partial r\neq 0$.

\subsubsection{Orbital Effects}
\label{sec:Orbital_Effects}
In their seminal work, \citet{Youdin2007} studied the diffusion of particles in Keplerian gas disks subjected to isotropic and homogeneous turbulence. They showed that orbital effects led to a decline in the strength of particle diffusion for large particles ($St \gtrsim 1$) with increasing Stokes number. Consequently, they revised the Schmidt number (Eq.~\ref{eq:def_Schmidt_number}) for diffusion in disks as follows:
\begin{equation}
    Sc^\mathrm{YL} \sim 1+St^2
\end{equation}
and also the radial particle diffusion coefficient
\begin{equation}\label{eq:YL_radial_diff}
    D_{d,r}^\mathrm{YL} \sim \frac{D}{1+St^2}
\end{equation}
where $D$ is the diffusion coefficient parametrized by the product of turbulent velocity dispersion squared and correlation time (Eq.~\ref{eq:charact_relation_D_t_corr}).\newline
Such orbital effects are not captured by the gradient diffusion model and must be parametrized explicitly. \newline
We highlight that the presence of, for example, a planet in a protoplanetary disk can introduce complexity to the orbital effects, potentially rendering an explicit parametrization inaccurate.\newline
After illustrating the limitations of the classical gradient diffusion model, we introduce a formalism that enables the derivation of improved turbulent transport models in the following section. 

\subsection{Reynolds-Averaged Mean-Flow Equations}
\label{eq:Mean-Flow Equations}
\label{sec:mean-flow_Reynolds}
The \textit{Reynolds-averaged Navier-Stokes} \citep[RANS, ][]{Reynolds1895} approach is a widely used technique in fluid dynamics to model turbulent gas, focusing on large-scale average behavior of hydrodynamic quantities rather than their instantaneous values on small scales.\newline
Our discussion follows the work of \cite{Cuzzi1993}, applying the RANS technique to the mass (Eq.~\ref{eq:mass_cons}) and momentum (Eq.~\ref{eq:mom_cons}) conservation equations that govern dust particle dynamics in protoplanetary disks. The resulting system of mean-flow equations describes the particle dynamics in a turbulent environment. However, without additional modeling, the system is not closed, meaning that the total number of independent variables exceeds the total number of independent equations, and explicit closure models are required.\newline 
We focus on the statistically averaged behavior of the instantaneous particle density $\rho_d(x_j,t)$ and the velocity $v_{i}(x_j,t)$, which depend on the spatial variables $x_{j=1,2,3}$ and time $t$. We decompose these variables into averaged and fluctuating components:
\begin{equation}
    \rho_d = \bar{\rho}_d+\rho_d'
\end{equation}
 \begin{equation}\label{eq:Reynolds_vel_def}
    v_{i} = \bar{v}_i+v_{i}'
\end{equation}
Here, the overbar $\bar{\:}$ denotes the Reynolds average and the prime ${\:}'$ denotes short-term fluctuations. This decomposition is possible as long as the characteristic length scales and timescales of fluctuations are small compared to those of the mean values. \newline
The averages of the fluctuating components vanish:
\begin{equation}\label{eq:fluctuations_vanish}
    \overline{\rho'_d}=0, \quad  \overline{v'_{i}}=0
\end{equation}
We interpret the average here primarily as a statistical ensemble average, although it can be equivalent to the time average under the ergodic hypothesis. \newline
Next, we apply the Reynolds decomposition to the instantaneous mass conservation equation (Eq.~\ref{eq:mass_cons}), decomposing both density and velocity into mean and fluctuating components:
\begin{equation} \label{eq:mass_cons_reynolds}
    \frac{\partial \bar{\rho}_d}{\partial t}+\frac{\partial \rho'_d}{\partial t}+\frac{\partial}{\partial x_j}\big(\bar{\rho}_d\bar{v}_{j}+\bar{\rho}_dv'_j+{\rho'}_d\bar{v}_{j}+{\rho}'_dv'_j\big)=0
\end{equation}
The averaging operator commutes with time and space derivatives, and already averaged quantities are considered constant. After applying the averaging operator to Eq.~\ref{eq:mass_cons_reynolds}, the equation becomes:
\begin{equation} \label{eq:mass_cons_reynolds_avgd}
    \frac{\partial \bar{\rho}_d}{\partial t}+\frac{\partial}{\partial x_j}\Big(\bar{\rho}_d\bar{v}_j+\overline{\rho'_d v'_{j}}\Big)=0
\end{equation}
This equation describes the dynamics of the mean particle density $\bar{\rho}_d$. Besides a \textit{mean advection flux} $\bar{\rho}_d\bar{v}_j$, the equation contains a new quantity, the \textit{mean turbulent mass flux} $\overline{\rho'_d v'_{j}}$, which can be interpreted as a mean particle mass flux driven by turbulence. The explicit form of this correlation term is unknown without further modeling, a condition known as the \textit{closure problem} \citep[see e.g.][for more details on the closure problem]{Fox03}.  \newline
The two flux components in Eq.~\ref{eq:mass_cons_reynolds_avgd} have independent dynamics, requiring additional equations to describe their evolution. To find these equations, we apply the Reynolds decomposition and averaging procedure to the momentum conservation equation (Eq.~\ref{eq:mom_cons}), yet ignoring gravity for simplicity, yielding:
\begin{equation}\label{eq:momentum_cons_reynolds_avgd}
\begin{split}
    \frac{\partial}{\partial t}\big(\bar{\rho}_d\bar{v}_i\big)+& \frac{\partial}{\partial t}\Big(\overline{\rho_d' v'_{i}}\Big)+\frac{\partial}{\partial x_j}\Big(
    \bar{\rho}_d\bar{v}_i\bar{v}_j+
    \overline{\rho'_dv'_{i}}\bar{v}_j+
    \bar{v}_i\overline{\rho'_dv'_{j}}+\\
    \underbrace{\overline{\rho}_d\overline{v'_{i} v'_{j}}}_\textrm{i}+
    &\underbrace{\overline{\rho'_dv'_{i} v'_{j}}}_\textrm{ii}\Big)=\bar{\rho}_d\frac{\bar{u}_i-\bar{v}_i}{t_s}+\underbrace{\frac{\overline{\rho'_d u'_{i}}-\overline{\rho'_d v'_{i}}}{t_s}}_\textrm{iii}
\end{split}
\end{equation}
Besides the mean turbulent mass flux $\overline{\rho'_d v'_{i}}$, three more terms (i, ii, and iii) contain unknown correlations. In the following section, we illustrate how a gradient diffusion approach can be used to close the system of Reynolds averaged mean flow equations via the approach of \citet{Huang22}.

\subsection{Recent Work}
\label{sec:recent_work}
\label{sec:GD_model}
The primary issue with the gradient diffusion model, as discussed in Sec.~\ref{sec:The Turbulent Particle Diffusion Model}, in the context of disk modeling, is probably its failure to conserve angular momentum \citep{Tominaga19}. Recently, two solutions to this problem have been proposed. We will briefly summarize these below.

\subsubsection{The Approach of \citet{Tominaga19} and \citet{Huang22}}
The first approach, proposed by \citet{Tominaga19} and further refined by \citet{Huang22}, is based on the Reynolds decomposition formalism by \citet{Cuzzi1993}. They argue that the term~$\mathrm{iii}$ in Eq.~\ref{eq:momentum_cons_reynolds_avgd} vanishes for small, well-coupled particles ($St\ll 1$). Similarly, the triple correlation term $\mathrm{ii}$ is typically argued to vanish as long as turbulent fluctuations are small \citep[e.g.][]{Blackman03}. The term~$\mathrm{i}$ in Eq.~\ref{eq:momentum_cons_reynolds_avgd} represents turbulent particle stresses, analogous to the Reynolds stress in the gas. The on-diagonal elements of the term~$\mathrm{i}$ represent the effect of a \textit{particle pressure}, similar to a thermal pressure in the gas \citep{Dobrovolskis99}.\newline
\cite{Shariff11} and \cite{Tominaga19}, we here use a closure relation of the form 
\begin{equation}\label{eq:turb_vel_dispersion}
    \overline{v'_{i} v'_{j}}=\delta_{ij}c_d^2
\end{equation}
to express the turbulent Reynolds stress in terms of a scalar particle velocity dispersion $c_d$. Here $\delta_{ij}$ is the Kronecker delta. For small particles, \cite{Huang22} argue the squared dispersion $c_d^2$ vanishes based on an argument by \cite{Garaud2004}, implying that all three terms, $\mathrm{i}$, $\mathrm{ii}$ and $\mathrm{iii}$ in Eq.~\ref{eq:momentum_cons_reynolds_avgd} can be neglected for tightly coupled particles. The mean momentum equation then reads: 
\begin{equation}\label{eq:momentum_cons_reynolds_avgd_2}
\begin{split}
    \frac{\partial}{\partial t}\big(\bar{\rho}_d\bar{v}_i\big)+&\underbrace{\frac{\partial}{\partial t}\Big(\overline{\rho_d' v'_{i}}\Big)}_\mathrm{I}+\frac{\partial}{\partial x_j}\Big(\bar{\rho}_d\bar{v}_i\bar{v}_j+
    \underbrace{
    \overline{\rho'_dv'_{i}}\bar{v}_j+
    \bar{v}_i\overline{\rho'_dv'_{j}}}_\mathrm{II}
    \Big)=\\ 
    &\frac{\bar{\rho}_d}{t_s}(\bar{u}_i-\bar{v}_i)
\end{split}
\end{equation}
Except for terms I and II, Eq.~\ref{eq:momentum_cons_reynolds_avgd_2} is equivalent to the instantaneous momentum equation (Eq.~\ref{eq:mom_cons}), with instantaneous variables replaced by their averages. \cite{Cuzzi1993} and \cite{Tominaga19} further neglect the term~I assuming it is small compared to the term to its left, yet offer no argument for this assumption. \cite{Huang22}, however, point out that the removal of term~I in Eq.~\ref{eq:momentum_cons_reynolds_avgd_2} would violate Galilean invariance, and therefore the term should be kept. In their work, \cite{Huang22} call the combined contribution of terms~I and II the \textit{momentum correction} that arises as a result of turbulent particle transport. \newline
The remaining unknown correlation in Eq.~\ref{eq:momentum_cons_reynolds_avgd_2} is $\overline{\rho_d' v'_{i}}$, generally does not vanish and therefore requires a closure relation. The simplest approach employs a \textit{gradient diffusion hypothesis} (GDH) which assumes turbulent mass flux is proportional to the gradient of the mean particle density \citep[see e.g.][]{Cuzzi1993,Tominaga19,Huang22}:
\begin{equation}\label{eq:GDH}
    \overline{\rho'_d v'_{i}} = -D\frac{\partial}{\partial x_i} \bar{\rho}_d
\end{equation}
As a result, the set of Reynolds averaged mean-flow equations is closed. Ignoring the momentum corrections (terms I and II), the equations recover the classical gradient diffusion model (Sec.~\ref{sec:The Turbulent Particle Diffusion Model}). \newline
The presented extension to the classical gradient diffusion model, based on robust mathematical foundations like Reynolds averages, indeed conserves total angular momentum \citep{Tominaga19, Huang22}. However, it does not resolve the other complications inherent to the gradient diffusion closure in Eq.~\ref{eq:GDH}. \newline
For illustrative purposes, we employ Eq.~\ref{eq:GDH} and Eq.~\ref{eq:mass_cons_reynolds_avgd}, rewriting the time derivative in term I of Eq.~\ref{eq:momentum_cons_reynolds_avgd_2} using averaged quantities: 
\begin{equation}\label{eq:xyz}
    \frac{\partial}{\partial t}\Big(\overline{\rho_d' v'_{i}}\Big)=D\frac{\partial}{\partial x_j}\frac{\partial}{\partial x_i}\bar{\rho}_d\bar{v}_j-D^2\frac{\partial^2}{\partial x^2_j}\frac{\partial}{\partial x_i}\bar{\rho}_d
\end{equation}
Assuming the diffusion coefficient $D$ to be constant, it can be moved inside the spatial derivatives and the entire r.h.s. of Eq.~\ref{eq:xyz} becomes a divergence, confirming the conservation of mean particle momentum in Eq.~\ref{eq:momentum_cons_reynolds_avgd_2}. However, the second term on the r.h.s. of Eq.~\ref{eq:xyz} may introduce nonphysical accelerations, an issue illustrated via the following one-dimensional example.\newline
Consider a static mean advection flow ($\overline{v}=\bar{u}=0$) with a small harmonic perturbation atop a constant particle density background $\bar{\rho}_d(x)=\rho_{d,0}(1+A\sin(kx))$, where $A\ll1$ and $k^{-1}$ characterizes the perturbation's length scale. Substituting this into Eq.~\ref{eq:xyz}, and subsequently into Eq.~\ref{eq:momentum_cons_reynolds_avgd_2}, the force term acting on the particle fluid is inversely proportional to the lengths scale of the perturbation to the third power:
\begin{equation}
    \frac{\partial}{\partial t}\big(\bar{\rho}_d\overline{v}\big)\propto k^3
\end{equation}
Consequently the diffusion time (Eq.~\ref{eq:diffusion timescale}) scales as $t_\mathrm{diff}\propto k^{-2}$, as expected of a diffusive solution. However, for small-scale perturbations ($k\to \infty$), the force acting on the dust fluid becomes arbitrarily large and thus the diffusion timescale $t_\mathrm{diff}$ arbitrarily small. This is inconsistent with the physical reality that the dust fluid can react to gas turbulence only on timescales similar to or larger than the stopping time $t_s$. Therefore, the gradient diffusion closure (Eq.~\ref{eq:GDH}) proves to be non-physical on small scales, smoothing out perturbations too quickly\footnote{Further complications arise when solving the mean flow equations numerically. We have found the third order spatial derivatives of the dust density, in Eq.~\ref{eq:xyz}, to be challenging to accurately compute when scales in the dust density become comparable to the computational grid.}.\newline 
This consideration can become important, e.g., when studying planetesimal formation via the gravitational collapse of small-scale particle overdensities. For instance, \citet{Umurhan20} showed that gradient diffusion suppresses the smallest modes of the streaming instability. In Sec.~\ref{sec:LPA}, we will discuss deviations from the strictly diffusive behavior at small scales that resolve this issue.\newline

\subsubsection{The Approach of \citet{Klahr2021}}
\citet{Klahr2021} do not employ Reynolds averages or the gradient diffusion hypothesis. Instead, they assume a settling-diffusion equilibrium ansatz, similar to the derivation of Brownian motion by \cite{Einstein1905}. Their dynamical equations read:
\begin{equation} \label{eq:Klahr_mass_cons}
    \frac{\partial \rho_d}{\partial t}+\frac{\partial}{\partial x_j}\big(\rho_d v_{j}\big)=0
\end{equation}
\begin{equation}\label{eq:Klahr_mom_cons}
        \frac{\partial}{\partial t}(\rho_dv_{i})+\frac{\partial}{\partial x_j}\bigg(\rho_dv_{i} v_{j}+\frac{1}{3}\frac{D}{t_s}\rho_d\delta_{ij}\bigg)=-\frac{\rho_d}{t_s}\big(v_{i}-u_{i}\big)+\rho_d g_i
\end{equation}
In these equations, the dust velocity $v_{i}$ represents the sum of the advection and diffusion velocities. Interestingly, these equations do not contain an explicit diffusion term, instead turbulent transport is modeled via a pressure-like term in the momentum equation. The equations conserve angular momentum and are significantly simpler than the previous model. \newline
As we will demonstrate in Sec.~\ref{sec:LPA}, the characteristic turbulent transport timescale in this model is limited from below by the stopping time~$t_s$. However, in the small-particle limit ($t_s \to 0$), the particle pressure diverges and the expression must be modified \citep[see Sec.~\ref{sec:Large Grains Limit} and ][for a more detailed discussion]{Klahr2021}.

\subsection{The Hinze-Tchen Model}
\label{sec:Hinze-Tchen Model}
In this section, we briefly review the \textit{Hinze-Tchen model}, that describes the mixing of particles embedded in turbulent gas in the absence of external forces (see e.g. \cite{Youdin2007} or \cite{Fan1998} for a more detailed review of the Hinze-Tchen model). \newline
The diffusion coefficient $D$ is defined as the time derivative of the mean squared displacement in the limit of $t\to \infty$ (Eq.~\ref{def:diffusion_coefficient}). Hinze and Tchen studied the time derivative of the mean squared displacement at arbitrary times $t$
\begin{equation}\label{eq:def_frak_D}
    \mathfrak{D}_g(t)= \int_0^t \mathrm{d}t'\:\overline{u'(t')u'(0)}
\end{equation}
where we define $\mathfrak{D}_g$ to be the time-dependent diffusion coefficient. \newline
In homogeneous and steady turbulence, the Hinze-Tchen model predicts the time-dependent diffusion coefficient in gas to read, 
\begin{equation}\label{eq:Hinze_tchen_gas_D}
    \mathfrak{D}_g(t)=D_g\big(1-e^{-t/t_\mathrm{corr}}\big)
\end{equation}
where $D_g$ is the diffusion coefficient in the limit $t\to \infty$  as defined in Eq.~\ref{def:diffusion_coefficient}. \newline
The motion of particles embedded in the turbulent fluctuations of the gas is described by a \textit{Langevin equation}
\begin{equation}
    \frac{\mathrm{d} v'}{\mathrm{d} t}=-\frac{v'-u'}{t_s}
\end{equation}
Assuming the particles are small compared to the smallest turbulence wavelength and are always trapped inside the same turbulent eddy, the turbulent particle velocity dispersion is related to the turbulent velocity dispersion of the gas as \citep{Fan1998}
\begin{equation}\label{eq:particle_vel_dispers}
    \overline{{v'}^2}=\frac{t_\mathrm{corr}}{t_\mathrm{corr}+t_s}\:\overline{{u'}^2}
\end{equation}
For small particles ($t_s\ll t_\mathrm{corr}$), the Hinze-Tchen model predicts the turbulent particle velocity dispersion to be equal to the dispersion in the gas $\overline{{v'}^2}\approx\overline{{u'}^2}$. For large particles ($t_s\gg t_\mathrm{corr}$), the turbulent particle dispersion scales inversely to the stopping time $\propto t_s^{-1}$. \newline 
We define the \textit{turbulence time} $t_t$ as
\begin{equation}\label{eq_turbulence_time}
    t_t \equiv t_\mathrm{corr}+t_s
\end{equation}
and combine Eq.~\ref{eq:particle_vel_dispers} with Eq.~\ref{eq:charact_relation_D_t_corr} to write the squared particle velocity dispersion as
\begin{equation}\label{eq:particle_vel_dipsersion}
    \overline{{v'}^2}=\frac{D_g}{t_t}
\end{equation}
Equivalently to Eq.~\ref{eq:Hinze_tchen_gas_D}, the Hinze-Tchen model also predicts a time dependent diffusion coefficient $\mathfrak{D}_d(t)$ for particles embedded in the turbulent gas: 
\begin{equation}\label{eq:Hinze_tchen_particle_D}
    \mathfrak{D}_d(t)=\frac{D_g}{t_s^2-t_\mathrm{corr}^2}\bigg[t_s^2\Big(1-e^{-t/t_s}\Big)-t_\mathrm{corr}^2\Big(1-e^{-t/t_\mathrm{corr}}\Big)\bigg]
\end{equation}
The time-dependent particle diffusion coefficient approaches $D_g$ on timescales $t\gg \max(t_s,t_\mathrm{corr})$ after which it is identical to the diffusion coefficient in the gas. \newline
Interestingly, the Hinze-Tchen model reveals that although the turbulent particle dispersion $\overline{{v'}^2}$ relies on the particle-gas coupling via the stopping time $t_s$ (Eq.~\ref{eq:particle_vel_dipsersion}), the particle diffusion coefficient on long timescales (and in the absence of external forces) remains independent of the stopping time, and consequently, the particle size (Eq.~\ref{eq:Hinze_tchen_particle_D}). The latter is equal to the diffusion coefficient of the gas $\mathfrak{D}_d(t\to\infty)=D_g$. This outcome, albeit somewhat counterintuitive, signifies that large particles weakly coupled to turbulent gas fluctuations do not diffuse less efficiently than the gas.\newline
The physical reasoning for this result is as follows \citep{Youdin2007}: For small particles well-coupled to the turbulent gas fluctuation, the motion of dust is identical to the motion of the gas and the equality $D_d=D_g$ is straightforward. In contrast, large particles ($t_s\gg t_\mathrm{corr}$) show a decrease in their squared turbulent velocity dispersion $\overline{v'^2}$ with an increase in the stopping time $\overline{v'^2}\propto t_s^{-1}$. Simultaneously, the particle mean-free path $l_\mathrm{mfp}$ increases with the stopping time $t_s$ ($l_\mathrm{mfp}=v't_s$). As such, these effects cancel out, leading to $D_d=D_g$ for large particles as well.

\section{The Turbulent Particle Pressure Model}
\label{sec:TheTPPM} 
After having introduced the necessary theoretical background, we now transition to the main focus of this paper. In this section, we present the derivation of a novel Eulerian turbulent dust transport model, starting with the introduction of \textit{Favre averaged} mean flow equations of dust dynamics (Sec.~\ref{sec:Favre-Averaged Mean-Flow Equations}) and appropriate turbulence closures (Sec.~\ref{sec:Turbulence Closure}). \newline
The advantage of Favre averaging over Reynolds averaging is that it removes the turbulent flux term from the continuity equation and reduces the number of terms in the averaged momentum equation by a factor of two\footnote{In their work, \citet{Champney90} studied particle turbulent transport using Favre averages, but they considered the elimination of the turbulent flux a major drawback of this method. They argued that the eliminated term is the \textit{key feature of compressible two-phase flows}, further emphasizing that turbulence should cause mixing irrespective of a vanishing mean velocity. The authors abandoned the approach, adopting Reynolds averages instead. As far as we know, Favre averaging has not been used again to study turbulent particle transport in protoplanetary disks. Importantly, the method presented in this paper allows for turbulent transport and mixing, even in the absence of mean velocity.}.

\subsection{Favre-Averaged Mean-Flow Equations}
\label{sec:Favre-Averaged Mean-Flow Equations}
The Favre average is a density-weighted average \citep{Favre65}, and we denote it by a tilde $\tilde{\:}\:$:
\begin{equation}\label{eq:prop_1}
    \tilde{v}_i \equiv \frac{ \overline{\rho_d v_{i}} }{\bar{\rho}_d}
\end{equation}
We define new fluctuations with respect to the Favre average
 \begin{equation}\label{eq:Favre_average_def}
    v_{i} = \tilde{v}_i+v_{i}''
\end{equation}
and note that the fluctuations $v_{i}''$ do not necessarily vanish when applying the averaging operator:
 \begin{equation}
    \overline{v''}_{i}\neq 0
\end{equation}
in contrast to the fluctuations with respect to the Reynolds average (see Eq.~\ref{eq:fluctuations_vanish}). Only the density-weighted fluctuations vanish under applying the averaging operator
\begin{equation}
    \overline{\rho_d v_{i}''}=0
\end{equation}
as can be seen by replacing $v_{i}''$ with Eq.~\ref{eq:Favre_average_def} and using the definition in Eq.~\ref{eq:prop_1}.\newline
Relating the Reynolds-averaged and Favre-averaged velocities using the aforementioned definitions, we obtain:
 \begin{equation}\label{eq:Rey_Favre_connection}
    \tilde{v}_i = \bar{v}_i+v^*_i
\end{equation}
where we have defined the \textit{turbulent transport velocity} as
\begin{equation}\label{eq:diff_velocity}  
    v^*_i = \frac{\overline{\rho'_d v'_{i}}}{\bar{\rho}_d}
\end{equation}
Next, we decompose the pressureless fluid equations and average them using the Favre decomposition, yielding a Favre-averaged continuity equation:
\begin{equation} \label{eq:favre_mass_eq} 
    \frac{\partial \bar{\rho}_d}{\partial t}+\frac{\partial}{\partial x_j}\big(\bar{\rho}_d\tilde{v}_j\big)=0
\end{equation}
Compared to the Reynolds averaging procedure, Favre averaging indeed eliminates the turbulent flux term from the mass conservation equation above. \newline
In the case of the momentum equation, we only decompose the velocity, not the density. For the interfluid term $\rho_d' u'_{i}$, we perform a Reynolds decomposition instead of a Favre decomposition, as a Favre average would be ill-defined. Making sure not to mix Favre averaged and Reynolds averaged quantities, the decomposed momentum equation reads
\begin{equation}\label{eq:Favre_decomp} 
\begin{split}
    &\frac{\partial}{\partial t}\big({\rho_d}\tilde{v}_i\big)+\frac{\partial }{\partial t}\big({\rho_d}v_{i}''\big)+\frac{\partial}{\partial x_j}\Big( {\rho_d}\tilde{v}_i \tilde{v}_j
    +\rho_d\tilde{v}_iv_{j}''+
    \rho_dv_{i}''\tilde{v}_j+\\
    &\rho_dv_{i}''v_{j}''\Big)= 
    \frac{1}{t_s}\Big(\bar{\rho}_d\bar{u}_i+\bar{\rho}_du_{i}'+\rho_d'\bar{u}_i+\rho_d'u_{i}'\Big)-
    \frac{1}{t_s}\Big(\rho_d \tilde{v}_i+\rho_d v_{i}''\Big)
\end{split}
\end{equation}
to which we then apply the averaging operator: 
\begin{equation}\label{eq:Favre_averaged_mom} 
    \frac{\partial}{\partial t}\big(\bar{\rho}_d\tilde{v}_i\big)+\frac{\partial}{\partial x_j}\Big( \bar{\rho}_d\tilde{v}_i \tilde{v}_j+\overline{\rho_dv_{i}''v_{j}''}\Big)=
    \frac{\bar{\rho}_d}{t_s}\big(\bar{u}_i-\tilde{v}_i\big)+\frac{1}{t_s}\overline{\rho_d'u_{i}'}
\end{equation}
In this equation, there is one unknown correlation containing dust quantities on the l.h.s. and one unknown interfluid correlation on the r.h.s., both requiring explicit modeling. It is now apparent that the Favre-averaged momentum equation (Eq.~\ref{eq:Favre_averaged_mom}) is much simpler than the Reynolds-averaged momentum equation (Eq.~\ref{eq:momentum_cons_reynolds_avgd}).

\subsection{Turbulence Closures}
\label{sec:Turbulence Closure}
A gradient diffusion closure, as discussed in Sec.~\ref{sec:GD_model}, cannot be employed for the Favre-averaged equations, because the turbulent mass flux does not appear explicitly in the mean-flow equations\citep{Champney90}. Consequently, the unknown correlations in Eq.~\ref{eq:Favre_averaged_mom} must be modeled explicitly.\newline
We first consider the \textit{turbulent pressure tensor} $P_{d,ij}\equiv\overline{\rho_dv_{i}''v_{j}''}$, a symmetric rank two tensor, containing correlations of density-weighted velocity fluctuations. Assuming isotropic and homogeneous turbulence, we follow \citet{Youdin2007} and \citet{Shariff11}, and assume the turbulent pressure tensor to be proportional to the identity matrix $P_{d,ij}=\frac{1}{3} \overline{\rho_dv_{i}''^2}\delta_{ij}$. \newline
Using the definitions from Eqs.~\ref{eq:Reynolds_vel_def}, \ref{eq:Favre_average_def} and \ref{eq:diff_velocity}, we rewrite the on-diagonal elements of the turbulent pressure tensor as the sum of three terms
\begin{equation}\label{eq:expandion_trubulent_pressure}
    \overline{\rho_dv_{i}''^2} = \bar{\rho}_d \overline{v_{i}'^2}-\bar{\rho}_d {v^{*2}_i}+\overline{\rho_d'v_{i}'^2}
\end{equation}
The third term on the r.h.s. of Eq.~\ref{eq:expandion_trubulent_pressure} is a triple correlation term, which, based on prevalent arguments in fluid dynamics, is either small or vanishes entirely \citep[see e.g.][for an overview of these arguments]{Blackman03}. For instance, one could apply Gaussian statistics to show that correlations of odd numbers vanish \citep[as elaborated by][]{Lesieur97}.\newline
The second term on the r.h.s. of Eq.~\ref{eq:expandion_trubulent_pressure} is generally smaller than the first term because turbulent transport, driven by turbulent velocity dispersion, cannot exceed the velocity dispersion itself, namely $v^{*2}_i\ll\overline{v_{i}'^2}$. As such, the term is also negligible. Consequently, the on-diagonal elements of the turbulent pressure tensor are well-approximated by the product of mean particle density and the time-averaged velocity fluctuation squared:
\begin{equation}
    P_{d,ii} \approx \frac{1}{3}\bar{\rho}_d \overline{v_{i}'^2}
\end{equation}
We then apply the Hinze-Tchen model \citep{Hinze59,Tchen47}, and specifically use Eq.~\ref{eq:particle_vel_dipsersion}, to arrive at the explicit closure relation
\begin{equation}\label{eq:closure_xy}
    \overline{\rho_dv_{i}''^2}=\bar{\rho}_d\frac{D}{t_t}
\end{equation}
in which the turbulent pressure tensor is proportional to the ratio of the diffusion coefficient $D$ and the turbulence time $t_t$. \newline
An important caveat is that the velocity $v_i''$ in Eq.~\ref{eq:closure_xy} is measured in a fixed Eulerian frame, while the Hinze-Tchen formalism is formulated in terms of Lagrangian velocities. Therefore, the equality in Eq.~\ref{eq:closure_xy} holds strictly only when the distance travelled by a particle during a correlation time $t_\mathrm{corr}$ is small compared to the scale of the system itself because at this point Eulerian and Lagrangian statistics are equivalent \citep{Biferale95}. Fortunately, this requirement is already implicitly satisfied by performing the Reynolds/Favre decomposition (see Sec.~\ref{eq:Mean-Flow Equations}).\newline
Next, we define the squared turbulent particle velocity dispersion explicitly as 
\begin{equation} \label{eq:turb_vel_disp_def}
    c_d^2 \equiv \frac{D}{t_t}
\end{equation}
and focus on the interfluid correlation $\overline{\rho'_d u'_{i}}$ that appears in the Favre-averaged momentum equation (Eq.~\ref{eq:Favre_averaged_mom}). In the short correlation time limit ($t_\mathrm{corr}\ll t_s$), particles are loosely coupled to turbulent fluctuations in the gas, and many turbulent eddies pass over an individual particle within one stopping time $t_s$. Consequently, in this limit, we expect particle fluid fluctuations to be entirely uncorrelated to turbulent fluctuations in the gas. Thus, the averaging operator in the second-order interfluid correlations acts independently on each quantity, i.e.
\begin{equation}\label{eq:interfluid_correlations}
    \overline{\rho'_d u'_{i}}=\overline{\rho'_d}\cdot\overline{ u'_{i}}
\end{equation}
where the r.h.s. is by definition zero (Eq.~\ref{eq:fluctuations_vanish}). \newline
In the short stopping time limit ($t_s\ll t_\mathrm{corr}$), the interfluid correlation does not vanish, and we expect particles to perfectly couple to the turbulent fluctuations in the gas. Therefore, in this limit, the relation $v'_{i}=u'_{i}$ holds \citep{Cuzzi1993}, yielding $\overline{\rho'_d u'_{i}}=\bar{\rho}_dv^*_i$. \newline 
We propose a simple closure relation, linearly connecting the two asymptotic cases of the interfluid correlation:
\begin{equation}\label{eq:dust_gas_vel_closure}
    \overline{\rho_d' u_{i}'}=\frac{t_\mathrm{corr}}{t_t}\: \bar{\rho}_d v^*_i
\end{equation}
The closure relation in Eq.~\ref{eq:dust_gas_vel_closure} has the expected asymptotic properties, namely, $\overline{\rho_d' u_{i}'}\simeq 0$ for $t_\mathrm{corr}\ll t_s$ and $\overline{\rho_d' u_{i}'}\simeq\bar{\rho}_d v^*_i$ for $t_s\ll t_\mathrm{corr}$. \newline
The set of mass and momentum equations using the new closure relations (given by Eq.~\ref{eq:closure_xy} and Eq.~\ref{eq:dust_gas_vel_closure}) read:
\begin{equation} \tag{\ref{eq:favre_mass_eq}}
    \frac{\partial \bar{\rho}_d}{\partial t}+\frac{\partial}{\partial x_j}\big(\bar{\rho}_d\tilde{v}_i\big)=0
\end{equation}
\begin{equation}\label{eq:Favre_mom_eq}
    \frac{\partial}{\partial t}\big(\bar{\rho}_d\tilde{v}_i\big)+\frac{\partial}{\partial x_j}\bigg( \bar{\rho}_d\tilde{v}_i \tilde{v}_j+\frac{1}{3}\bar{\rho}_dc_d^2\delta_{ij}\bigg)=-
    \frac{1}{t_s}\bar{\rho}_d\big(\bar{v}_i-\bar{u}_i\big)-
    \frac{1}{t_t} \bar{\rho}_dv^*_i
\end{equation}
In the system above, Eq.~\ref{eq:favre_mass_eq} represents the mass-conservation equation without source terms, indicating that a local change in mean density $\bar{\rho}_d$ is only governed by the divergence of the total mass flux $\bar{\rho}_d\tilde{v}_i$. The dynamics of the total mass flux are described by Eq.~\ref{eq:Favre_mom_eq}, containing the turbulent pressure term $\bar{\rho}_d c_d^2$ that drives the dust momentum transport. \newline
The right-hand side of Eq.~\ref{eq:Favre_mom_eq} contains two dissipative terms. The first arises from aerodynamic drag between the particles and the gas fluid, acting on a timescale equal to the stopping time $t_s$. Note how the drag term acts on the mean velocities $\bar{v}_i$ and $\bar{u}_i$, respectively, and not on the Favre-averaged velocities ($\tilde{v}_i$, $\tilde{u}_i$). The second dissipative term dampens the transport effects caused by the turbulent particle velocity dispersion on a timescale $t_t$, which is generally longer than the stopping time ($t_t \geq t_s$). Only when $t_s\gg t_\mathrm{corr}$, the timescales are almost identical ($t_t\simeq t_\mathrm{corr}$). On timescales significantly shorter than $t_t$, the turbulent pressure term is the only relevant term for particle dynamics, a regime known as the \textit{ballistic} regime \citep[][in contrast to the \textit{diffusive} regime on long timescales]{Taylor20}. \newline
It is crucial to note that the system of Eq.~\ref{eq:favre_mass_eq} and Eq.~\ref{eq:Favre_mom_eq} includes three different mean particle velocities, the ensemble-averaged velocity $\bar{v}_i$, the turbulent transport velocity $v^*_i$, and the density weighted mean velocity $\tilde{v}_i$. While these velocities are related via Eq.~\ref{eq:Rey_Favre_connection}, there are more variables than equations and the system in its most general form is not yet closed. We will address this issue in the following section.

\subsection{Three-Equation Formalism}
\label{sec:Three-Equation Model}
After highlighting that the system in the form of Eq.~\ref{eq:favre_mass_eq} and Eq.~\ref{eq:Favre_mom_eq} is not yet closed, we address this issue in this section. \newline
We focus on the momentum equation and use Eq.~\ref{eq:Rey_Favre_connection} to rewrite the first term of Eq.~\ref{eq:Favre_mom_eq}
\begin{equation}
     \frac{\partial}{\partial t}\big(\bar{\rho}_d\tilde{v}_i\big) =  \frac{\partial}{\partial t}\big(\bar{\rho}_d\bar{v}_i\big) +  \frac{\partial}{\partial t}\big(\bar{\rho}_dv^*_i\big)
\end{equation}
and recognize that the local rate of change of the total momentum $\bar{\rho}_d\tilde{v}_i$ can be written as the sum of the local rate of change of the mean momentum $\bar{\rho}_d\bar{v}_i$ and the mean turbulent flux $\bar{\rho}_dv^*_i$. Motivated by this insight, we aim to decompose Eq.~\ref{eq:Favre_mom_eq} into two separate (but coupled) momentum equations, each describing the dynamics of either the mean momentum $\bar{\rho}_d\bar{v}_i$ or the mean turbulent flux $\bar{\rho}_dv^*_i$. In other words, we aim to find two equations
\begin{equation}\label{eq:pseudo_eq_1}
    \frac{\partial}{\partial t}\big(\bar{\rho}_d\bar{v}_i\big) = ...
\end{equation}
and 
\begin{equation}\label{eq:pseudo_eq_2}
    \frac{\partial}{\partial t}\big(\bar{\rho}_dv^*_i\big) = ...
\end{equation}
such that the sum of Eq.~\ref{eq:pseudo_eq_1} and Eq.~\ref{eq:pseudo_eq_2} equals Eq.~\ref{eq:Favre_mom_eq}.\newline
To find these equations, we consider two special cases. The first case is the limit of vanishing turbulence ($D\to 0$). In this limit, $v_i^*=0$ holds, thus $\partial\bar{\rho}_dv^*_i /\partial t = 0$, and Eq.~\ref{eq:Favre_mom_eq} can be written as 
\begin{equation}\label{eq:momentum_like_ax2} 
    \frac{\partial}{\partial t}\big(\bar{\rho}_d\bar{v}_i\big)+\frac{\partial}{\partial x_j}\big( \bar{\rho}_d\bar{v}_i\bar{v}_j\big)=
    -\frac{1}{t_s}\bar{\rho}_d\big(\bar{v}_i-\bar{u}_i\big)
\end{equation}
Thus, we find an expression for Eq.~\ref{eq:pseudo_eq_1} in this special limit. \newline
We compare this to a second case in which turbulence is present ($D\neq 0$) and the mean gas velocity is zero ($\bar{u}=0$). Additionally, we assume the stopping time to be short ($t_s\ll t$) such that in this case $\bar{v}_i=0$ holds, thus $\partial\bar{\rho}_d \bar{v}_i /\partial t = 0$, and Eq.~\ref{eq:Favre_mom_eq} simplifies to 
\begin{equation}\label{eq:momentum_like_bx2} 
    \frac{\partial}{\partial t}\big(\bar{\rho}_dv^*_i\big)+\frac{\partial}{\partial x_j}\bigg( \bar{\rho}_dv^*_i v^*_j+\delta_{ij}\frac{1}{3}\bar{\rho}_d c_d^2\bigg)=-\frac{1}{t_t}\bar{\rho}_dv^*_i
\end{equation}
Thus, we have found the functional form of Eq.~\ref{eq:pseudo_eq_2} in this special case.  \newline
Next, we aim to find the appropriate expressions for a general case. As mentioned above, we also require the sum of Eq.~\ref{eq:pseudo_eq_1} and Eq.~\ref{eq:pseudo_eq_2} to equal Eq.~\ref{eq:Favre_mom_eq} for a general case. We note that the sum of Eq.~\ref{eq:momentum_like_ax2} and Eq.~\ref{eq:momentum_like_bx2} does not equal Eq.~\ref{eq:Favre_mom_eq} but is missing two terms: $\partial/\partial x_j(\bar{\rho}_d\bar{v}_i v^*_j)$ and $\partial/\partial x_j(\bar{\rho}_dv^*_i \bar{v}_j)$.
These terms vanish in the two aforementioned special cases, thus, we cannot yet assign them unambiguously to either Eq.~\ref{eq:momentum_like_ax2} or Eq.~\ref{eq:momentum_like_bx2}. \newline
However, we argue that the only possibility to ensure that the sum of Eq.~\ref{eq:momentum_like_ax2} and Eq.~\ref{eq:momentum_like_bx2} is equal to Eq.~\ref{eq:Favre_mom_eq}, while at the same time ensuring Galilean invariance and momentum conservation of each equation individually, is to assign the term $\partial/\partial x_j(\bar{\rho}_d\bar{v}_i v^*_j)$ to Eq.~\ref{eq:momentum_like_ax2}, and to assign term $\partial/\partial x_j(\bar{\rho}_dv^*_i \bar{v}_j)$ to Eq.~\ref{eq:momentum_like_bx2}.\newline 
Consequently, the general form of the two decomposed momentum equations is the following:  
\begin{equation}\label{eq:momentum_like_a} 
    \frac{\partial}{\partial t}\big(\bar{\rho}_d\bar{v}_i\big)+\frac{\partial}{\partial x_j}\Big( \bar{\rho}_d\bar{v}_i\cdot \big(\bar{v}_j+v^*_j\big)\Big)=
    -\frac{1}{t_s}\bar{\rho}_d\big(\bar{v}_i-\bar{u}_i\big)
\end{equation}
\begin{equation}\label{eq:momentum_like_b} 
    \frac{\partial}{\partial t}\big(\bar{\rho}_dv^*_i\big)+\frac{\partial}{\partial x_j}\bigg( \bar{\rho}_dv^*_i\cdot \big(\bar{v}_j+v^*_j\big)+\delta_{ij}\frac{1}{3}\bar{\rho}_d c_d^2\bigg)=-\frac{1}{t_t}\bar{\rho}_dv^*_i
\end{equation}
It is straightforward to show that the sum of Eq.~\ref{eq:momentum_like_a} and Eq.~\ref{eq:momentum_like_b} is indeed equal to Eq.~\ref{eq:Favre_mom_eq}. {Moreover, both transport terms can be written as a divergence, and thus do not contribute to the non-conservation of momentum. Note, Eq.~\ref{eq:momentum_like_a} and Eq.~\ref{eq:momentum_like_b} both have a dissipative term on their right-hand side. Thus, neither equation by itself is momentum-conserving. When we discuss the turbulent gas equations in Sec.~\ref{sec:Including Turbulent Gas Dynamics}, we will show that both dissipative terms also appear in the gas momentum equations (with opposite sign), thus ensuring momentum conservation in the full system.} \newline
Including the continuity equation in the following form
\begin{equation} \label{eq:mass_eq_22}
    \frac{\partial \bar{\rho}_d}{\partial t}+\frac{\partial}{\partial x_j}\Big(\bar{\rho}_d\big(\bar{v}_j+v^*_j\big)\Big)=0,
\end{equation}
the system of Eq.~\ref{eq:momentum_like_a}, Eq.~\ref{eq:momentum_like_b} and Eq.~\ref{eq:mass_eq_22} now represents a closed system. \newline
In this \textit{three-equation formalism}, Eq.~\ref{eq:momentum_like_a} describes the dynamics of the mean particle momentum  $\bar{\rho}_d\bar{v}_i$. It contains the explicit drag term, and in the presence of external forces, would also contain a gravity term. In contrast, Eq.~\ref{eq:momentum_like_b} describes the dynamics of the turbulent mass flux $\bar{\rho}_dv^*_i$. It includes a turbulent pressure term that drives turbulent transport via a pressure gradient force with the turbulent speed $c_d$. The dissipative term on the r.h.s. of Eq.~\ref{eq:momentum_like_b} reestablishes equilibrium flow over a timescale $t_t$. \newline
It is now apparent that the turbulent particle pressure model, as formulated above, permits non-equilibrium turbulent particle transport via an additional transport equation (Eq.~\ref{eq:momentum_like_b}), in other words, it allows for non-local transport effects.\newline
In the following section, we will discuss the properties of the turbulent particle pressure model and compare it to the classical diffusion approaches.

\section{Applications}
\label{sec:Properties and Applications of the TPP Model}
After deriving a novel model for turbulent particle transport, we study its applications. First, to simple illustrative examples in Sec.~\ref{eq:Turbulent Particle Transport Beyond Diffusion}, and then to protoplanetary disks in Sec.~\ref{sec:Vertical Steady-State Disk Profile}. For this, we first need to explicitly consider the turbulent gas background, which we will discuss in Sec.~\ref{sec:Including Turbulent Gas Dynamics}. 

\subsection{Including Turbulent Gas Dynamics}
\label{sec:Including Turbulent Gas Dynamics}
In this discussion of gas dynamics, we consider a locally isothermal gas fluid in local thermodynamical equilibrium (LTE), modeled by the set of locally isothermal Euler equations, given by (in the absence of external forces)
\begin{equation} \label{eq:mass_cons_gas}
    \frac{\partial \rho_g}{\partial t}+\frac{\partial}{\partial x_j}\big(\rho_g u_{j}\big)=0
\end{equation}
\begin{equation}\label{eq:mom_cons_gas}
        \frac{\partial}{\partial t}(\rho_du_{i})+\frac{\partial}{\partial x_j}\big(\rho_gu_{i} v_{j}\big)+\frac{\partial}{\partial x_i}\big(\rho_g c_s^2\big)=-\frac{\rho_d}{t_s}\big(u_{i}-v_{i}\big)
\end{equation}
These equations are the continuity equation (Eq.~\ref{eq:mass_cons_gas}) and the momentum equations (Eq.~\ref{eq:mom_cons_gas}). They neglect the effects of molecular viscosity, which is a valid assumption for turbulent protoplanetary disks \citep[][]{Shu92}. The locally isothermal equation of state eliminates the need for an energy equation, simplifying the analysis. The term on the r.h.s of Eq.~\ref{eq:mom_cons_gas} describes the exchange of momentum with the dust via aerodynamic drag, i.e., the back reaction. \newline
Analogous to the procedure in Sec.~\ref{sec:Favre-Averaged Mean-Flow Equations}, we formulate a set of mean-flow equations from Eqs.~\ref{eq:mass_cons_gas} and \ref{eq:mom_cons_gas} using Favre-averages:
\begin{equation} \label{eq:favre_mass_eq_gas}
    \frac{\partial \bar{\rho}_g}{\partial t}+\frac{\partial}{\partial x_j}\big(\bar{\rho}_g\tilde{u}_i\big)=0
\end{equation}
\begin{equation}\label{eq:Favre_averaged_mom_gas} 
\begin{split}
    \frac{\partial}{\partial t}\big(\bar{\rho}_g\tilde{u}_i\big)+&\frac{\partial}{\partial x_j}\Big( \bar{\rho}_g\tilde{u}_i \tilde{u}_j+\underbrace{\overline{\rho_gu_{i}''u_{j}''}}_\mathrm{I}\Big)+\frac{\partial}{\partial x_i}\big(\rho_g c_s^2\big)=\\
    &-\frac{\bar{\rho}_d}{t_s}\Big(\bar{u}_i-\tilde{v}_i\Big)-\frac{1}{t_s}\underbrace{\overline{\rho_d'u_{i}'}}_\mathrm{II}
\end{split}
\end{equation}
Notably, the averaged continuity equation (Eq.~\ref{eq:favre_mass_eq_gas}) is formally equivalent to the instantaneous continuity equation (Eq.~\ref{eq:mass_cons_gas}) with the instantaneous variables replaced by the averaged variables ($\rho_g,u_i \to \bar{\rho}_g,\tilde{u}_i$). The source terms on the r.h.s. of Eq.~\ref{eq:Favre_averaged_mom_gas} are identical to the source term of Eq.~\ref{eq:Favre_averaged_mom} (multiplied by $-1$), indicating that, even though momentum is exchanged between the gas and the dust, momentum is globally conserved. \newline
Adopting the same closure approach as for dust, we use the closure relation of Eq.~\ref{eq:dust_gas_vel_closure} for term $\mathrm{II}$ in Eq.~\ref{eq:Favre_averaged_mom_gas} and recognize the turbulent pressure tensor $P_{g,ij}=\overline{\rho_g u_{i}''u_{j}''}$ in term $\mathrm{I}$. 
We further decompose the turbulent pressure tensor into a traceless and an isotropic tensor
\begin{equation}\label{eq:pressure_tensor}
    P_{g,ij} = R_{ij}+p_t\delta_{ij}
\end{equation}
Here, the \textit{isotropic turbulent pressure} $p_t$ is defined as 
\begin{equation}\label{eq:iso_turb_pressure}
    p_t = \overline{\rho_g u''^2}
\end{equation}
and the traceless Reynolds tensor as
\begin{equation}\label{eq:Reynolds_tensore}
    R_{ij} = \overline{\rho_g u_{i}''u_{j}''} - p_t\delta_{ij}.
\end{equation}
Incorporating these definitions, we rewrite the mean-flow gas momentum equation as follows:
\begin{equation}\label{eq:Favre_averaged_mom_gas_3} 
\begin{split}
    \frac{\partial}{\partial t}\big(\bar{\rho}_g\tilde{u}_i\big)+&\frac{\partial}{\partial x_j}\big( \bar{\rho}_g\tilde{u}_i \tilde{u}_j+R_{ij}\big)+\frac{\partial}{\partial x_i}\big(\bar{\rho}_g c_s^2+p_t\big)=\\
    &-\frac{\bar{\rho}_d}{t_s}\big(\bar{u}_i-\bar{v}_i\big)+\frac{1}{t_t}\bar{\rho}_dv^*_i
\end{split}
\end{equation}
Analogous to Sec.~\ref{sec:Turbulence Closure}, we could now use the Hinze-Tchen formalism and Eq.~\ref{eq:charact_relation_D_t_corr} to rewrite the isotropic turbulent pressure $p_t$ in terms of the diffusion coefficient $D$ and the correlation time $t_\mathrm{corr}$. However, assuming subsonic turbulence as prevalent in protoplanetary disks \citep{Hughes11,Guilloteau12,Flaherty15,Flaherty18,Teague16}, the turbulent pressure is vanishingly small compared to the thermal pressure $p_t \ll \bar{\rho}_g c_s^2$. Consequently, we safely neglect the isotropic turbulent pressure in Eq.~\ref{eq:Favre_averaged_mom_gas} ($\bar{\rho}_g c_s^2+p_t\approx \bar{\rho}_g c_s^2$). \newline
Next, we need to specify a turbulence model for the Reynolds tensor $R_{ij}$. Among the numerous models available in fluid dynamics literature, the eddy viscosity model, also known as the \textit{Boussinesq hypothesis} \citep[see e.g.][]{Champney90}, is frequently used in the protoplanetary disk community. This model treats turbulent stresses in gas as an effective turbulent viscosity. The viscous stress tensor in Cartesian coordinates reads \citep[e.g.][]{Shu92}
\begin{equation}\label{eq;viscous_stress_tensor}
   R_{ij} = \rho_g \nu \bigg(\frac{\partial u_{i}}{\partial x_j}+\frac{\partial u_{j}}{\partial x_i} - \frac{2}{3} \delta_{ij} \bf{\nabla\cdot u}\bigg)
\end{equation}
and is parametrized by the turbulent viscosity $\nu$, typically parametrized using the $\alpha$-description of  \citet{Shakura1973}:
\begin{equation}\label{eq:alpha_parametrization}
    \nu = \alpha c_s h_g
\end{equation}  
We then use the gradient diffusion hypothesis to close the last remaining unknown correlation term. Specifically, we write
\begin{equation}\label{eq:gas_closure_22}
    \overline{\rho'_g u'_{i}}=-D \frac{\partial}{\partial x_i} \rho_g
\end{equation}
such that we can use Eq.~\ref{eq:gas_closure_22} and the relation between the turbulent mass flux and the Favre average (analogous to Eq.~\ref{eq:Rey_Favre_connection} and Eq.~\ref{eq:diff_velocity} for dust), to write the mean gas velocity $\bar{u}_i$ as follows: 
\begin{equation}\label{eq:mean_gas_velc}
    \bar{u}_i=\tilde{u}_i - \frac{D}{\rho_g} \frac{\partial}{\partial x_i} \rho_g.
\end{equation}
Plugging Eq.~\ref{eq:mean_gas_velc} into Eq.~\ref{eq:Favre_averaged_mom_gas_3}, we rewrite the mean-flow equation of the gas momentum as follows:
\begin{equation}\label{eq:Favre_averaged_mom_gas_2} 
\begin{split}
    \frac{\partial}{\partial t}\big(\bar{\rho}_g\tilde{u}_i\big)+&\frac{\partial}{\partial x_j}\big( \bar{\rho}_g\tilde{u}_i \tilde{u}_j+\sigma_{ij}\big)+\frac{\partial}{\partial x_i}\big(\bar{\rho}_g c_s^2\big)=\\
    &-\frac{\bar{\rho}_d}{t_s}\big(\tilde{u}_i-\bar{v}_i\big)
    +\frac{1}{t_t}\bar{\rho}_dv^*_i - \frac{D}{t_s}\frac{\bar{\rho}_d}{\bar{\rho}_g} \frac{\partial}{\partial x_i} \bar{\rho}_g
\end{split}
\end{equation}
The above equation now only contains one explicit gas velocity variable ($\tilde{u}_i$). \newline
The full two-fluid system of equations (gas+dust) then reads
\begin{equation} \tag{\ref{eq:favre_mass_eq_gas}}
    \frac{\partial \bar{\rho}_g}{\partial t}+\frac{\partial}{\partial x_j}\big(\bar{\rho}_g\tilde{u}_i\big)=0
\end{equation}
\begin{equation}\tag{\ref{eq:Favre_averaged_mom_gas_2}}
\begin{split}
    \frac{\partial}{\partial t}\big(\bar{\rho}_g\tilde{u}_i\big)+\frac{\partial}{\partial x_j}\big( \bar{\rho}_g\tilde{u}_i \tilde{u}_j+\sigma_{ij}\big)+\frac{\partial}{\partial x_i}\big(\bar{\rho}_g c_s^2\big)=\\
    -\frac{\bar{\rho}_d}{t_s}\big(\tilde{u}_i-\bar{v}_i\big)
    +\frac{1}{t_t}\bar{\rho}_dv^*_i - \frac{D}{t_s}\frac{\bar{\rho}_d}{\bar{\rho}_g} \frac{\partial}{\partial x_i} \bar{\rho}_g
\end{split}
\end{equation}
\begin{equation} \tag{\ref{eq:mass_eq_22}}
    \frac{\partial \bar{\rho}_d}{\partial t}+\frac{\partial}{\partial x_j}\Big(\bar{\rho}_d\big(\bar{v}_j+v^*_j\big)\Big)=0
\end{equation}

\begin{subequations}
\begin{equation}\label{eq:momentum_like_aa}
\begin{split}
    \frac{\partial}{\partial t}\big(\bar{\rho}_d\bar{v}_i\big)+\frac{\partial}{\partial x_j}\Big( \bar{\rho}_d\bar{v}_i\cdot \big(\bar{v}_j+v^*_j\big)\Big)=&
    -\frac{1}{t_s}\bar{\rho}_d\big(\bar{v}_i-\tilde{u}_i\big)+\\ 
    &\frac{D}{t_s}\frac{\bar{\rho}_d}{\bar{\rho}_g} \frac{\partial}{\partial x_i} \bar{\rho}_g
\end{split}
\end{equation}
\begin{equation}\label{eq:momentum_like_bb}
    \frac{\partial}{\partial t}\big(\bar{\rho}_dv^*_i\big)+\frac{\partial}{\partial x_j}\bigg( \bar{\rho}_dv^*_i\cdot \big(\bar{v}_j+v^*_j\big)+\delta_{ij}\frac{1}{3}\bar{\rho}_d c_d^2\bigg)=-\frac{1}{t_t}\bar{\rho}_dv^*_i
\end{equation}
\end{subequations}
and consists of a total of eleven equations describing the dynamics of eleven unknowns ($\bar{\rho}_g,\bar{\rho}_d,\allowbreak \tilde{u}_{i=1,2,3},\tilde{v}_{i=1,2,3},v^*_{i=1,2,3}$) and, thus, is a closed system of equations. \newline
The equations for gas are the continuity equation (Eq.~\ref{eq:favre_mass_eq_gas}) and the momentum equations (Eq.~\ref{eq:Favre_averaged_mom_gas_2}) which now looks like the Navier-Stokes equation with three additional terms accounting for the interaction with the dust. The dust continuity equation (Eq.~\ref{eq:mass_eq_22}) contains two mass flux, the mean mass flux $\bar{\rho}_d\bar{v}_i$, and the turbulent mass flux $\bar{\rho}_dv^*_i$, that govern the local change of the mean gas density $\bar{\rho}_d$. The equation describing the dynamics of the mean mass flux $\bar{\rho}_d\bar{v}_i$ (Eq.~\ref{eq:momentum_like_aa}) looks like the pressureless momentum equation with which we started in Eq.~\ref{eq:mom_cons}, but with the instantaneous velocity $v_i$ replaced by the mean velocity $\bar{v}$. However, it contains a new transport term on the left-hand side and a new term accounting for turbulent flows of the gas on the right-hand side. From the transport term $\bar{\rho}_d\bar{v}_i\cdot \big(\bar{v}_j+v^*_j)$ in Eq.~\ref{eq:momentum_like_aa}, it becomes apparent that turbulence can transport mean momentum via the turbulent velocity component $v^*_j$, which is the key distinction compared to the classical gradient diffusion model. Lastly, Eq.~\ref{eq:momentum_like_bb} is a new additional momentum equation that describes the dynamics of the turbulent mass flux $\bar{\rho}_dv^*_i$. It contains the turbulent pressure term $\bar{\rho}_d c_d^2$ that drives the turbulent transport, and a term $-\bar{\rho}_dv^*_i/t_t$ that acts to dissipate any directed turbulent transport. Note, in Eq.~\ref{eq:momentum_like_aa}, we used Eq.~\ref{eq:mean_gas_velc} to rewrite the gas velocity in the drag term, which is different from Eq.~\ref{eq:momentum_like_a}. External forces, when present, would appear in Eq.~\ref{eq:Favre_averaged_mom_gas_2} and in Eq.~\ref{eq:momentum_like_aa}. \newline
The two-fluid system above is Galilean invariant, and it conserves total momentum (angular and linear) globally.

\subsection{Turbulent Particle Transport Beyond Diffusion}
\label{eq:Turbulent Particle Transport Beyond Diffusion} 
\begin{figure}
\centering
\includegraphics[width=1.0\columnwidth]{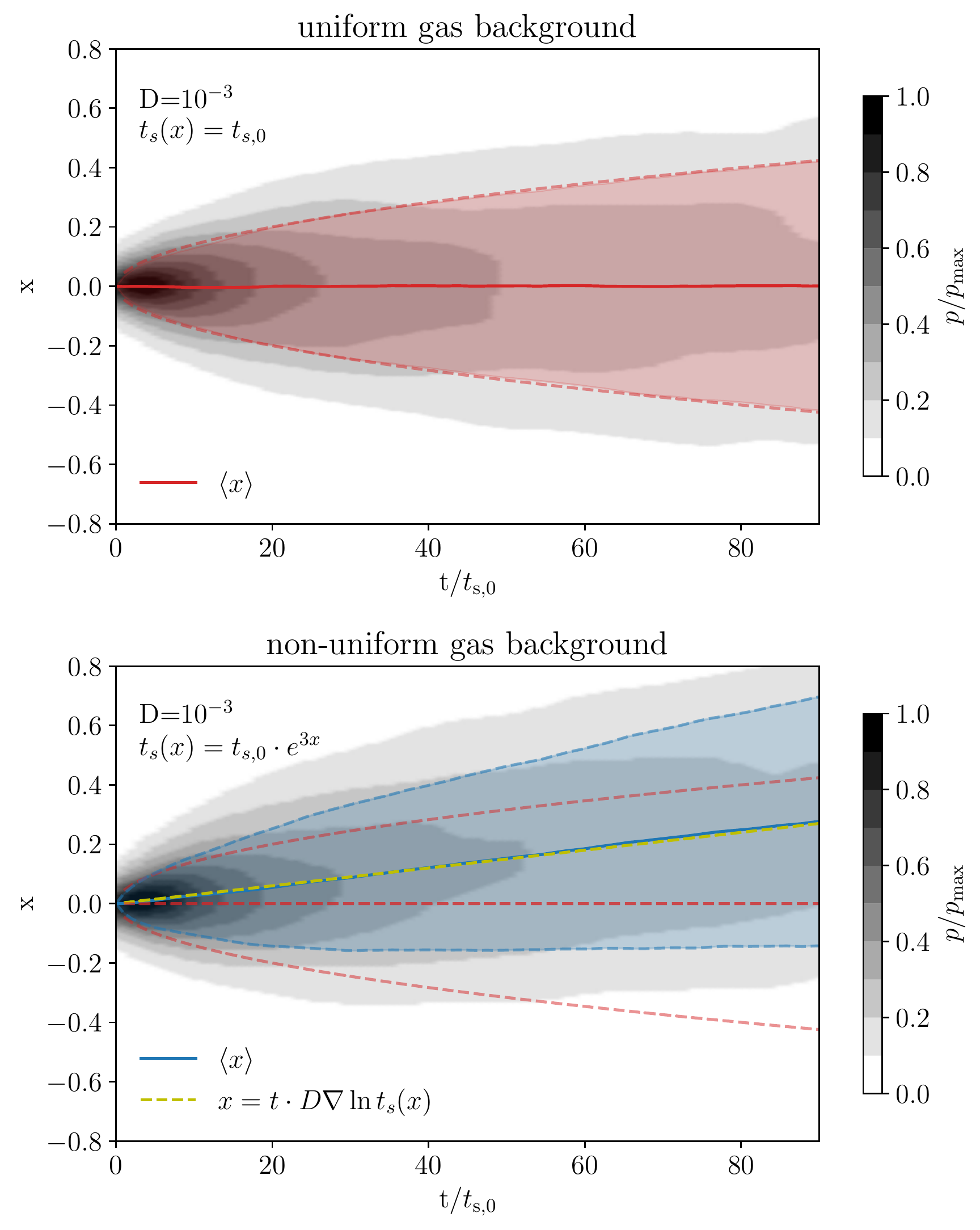}
\caption[Non-Fickian turbulent particle transport in non-uniform gas background.]{Space-time plot of the evolution of $N=2\cdot10^3$ individual (Lagrangian) particles, governed by the stochastic equation of motion (Eqs. \ref{eq:lagr_3}, \ref{eq:SEOM2}, \ref{eq:SEOM3}) in the absence of external forces. The gray background colors represent the normalized kernel density estimation of the particle distribution in $x$-$t$-space. All the particles are initially at rest at $x=0$. The solid lines show, for each time $t$, the mean of the distribution $\langle x \rangle$ , the colored shaded region covers a region within one standard deviation of the mean value. The correlation time is kept small compared to the stopping time ($t_\mathrm{corr}=0.01)$. \textit{Top:} Particles move through a uniform gas background such that the stopping time of the particles is constant in space $t_{s,0}=1$ (in arbitrary units). The diffusively spreading distribution remains centered around $x=0$. \textit{Bottom:} The particles move through a nonuniform gas background in which the stopping time $t_s$ increases exponentially in positive $x$-direction. As a result, the entire distribution drifts with a systematic and constant velocity $v=D\nabla \ln t_s$ towards increasing values of the stopping time (green dashed line) as predicted by Eq.~\ref{eq:steady_state_transport_flux}. The mean of the distribution $\langle x \rangle$ is shown with a solid blue line and follows the green dashed line. For a visual comparison, the mean and standard deviation of the example in a uniform background is plotted with red dashed lines also in the bottom subplot.}
\label{fig:random_walk}
\end{figure}

Before we discuss the system in the presence of gravity, we discuss a simple example in the absence of external forces that illustrates the main difference between the turbulent particle pressure model and gradient diffusion. \newline
We consider a dust distribution in a static gas background ($\tilde{u}_i=0$) and a small dust-to-gas ratio ($\rho_d/\rho_g \ll 1$) such that the static equilibrium in gas is not affected by the motion of the dust. Further, we consider a quasi-steady state of the dust in force balance, such that the source terms in Eq.~\ref{eq:momentum_like_aa} and in Eq.~\ref{eq:momentum_like_bb}  respectively cancel each other. \newline
The conditions for force balance in Eq.~\ref{eq:momentum_like_aa} is 
\begin{equation}\label{eq:mean_mass_flux_Reynolds}
    \bar{\rho}_d\bar{v}_i=D \frac{\bar{\rho}_d}{\bar{\rho}_g} \frac{\partial}{\partial x_i} \bar{\rho}_g
\end{equation}
This result illustrates that the dust couples to the advection flow of the gas such that we find a mean dust flow against the mean turbulent gas flow $\bar{v}_i=\bar{u}_i=-u^*_i$, where the last equality follows from $\tilde{u}_i=0$. \newline
The condition for force balance in Eq.~\ref{eq:momentum_like_bb} reads (assuming a constant diffusion coefficient D)
\begin{equation}\label{eq:turbulent_mass_flux_Favre}
    \bar{\rho}_d v^*_i = -D\frac{\partial}{\partial x_i}\bar{\rho}_d-D\bar{\rho}_d\frac{\partial}{\partial x_i}\ln t_t^{-1}
\end{equation}
The Eq.~\ref{eq:turbulent_mass_flux_Favre} describes the turbulent dust mass flux. In the first term on the r.h.s, we immediately recognize the gradient diffusion flux from Eq.~\ref{eq:GDH}. The second term on the r.h.s. is a novel contribution, predicting directed turbulent transport in the direction of increasing values of $t_t$ \footnote{Defining an entropy in the dust fluid as $s \equiv \ln{Dt_t^{-1}}$, the second term in Eq.~\ref{eq:turbulent_mass_flux_Favre} can also be interpreted transport across an entropy gradient.}. \newline
We use Eq.~\ref{eq:Rey_Favre_connection} to combine the equilibrium flux from Eq.~\ref{eq:mean_mass_flux_Reynolds} and that from Eq.~\ref{eq:turbulent_mass_flux_Favre} to arrive at the total dust mass flux:
\begin{equation}\label{eq:steady_state_transport_flux} 
    \bar{\rho}_d\tilde{v}_i=-D\rho_g\frac{\partial}{\partial x_i}\frac{\bar{\rho}_d}{\bar{\rho}_g}-D\bar{\rho}_d\frac{\partial}{\partial x_i}\ln t_t^{-1}
\end{equation}
Note, the first term on the r.h.s. now contains the gradient of the \textit{dust-to-gas ratio}. \newline
We conclude that the first term of the turbulent transport flux $\bar{\rho}_d v^*_i$ in Eq.~\ref{eq:turbulent_mass_flux_Favre} is given solely by the absolute gradient of the dust density. Thus, any derivation considering the particle distribution in isolation will arrive at this functional form of the turbulent transport term. However, in this quasi-steady state, the dust couples to the flow of the gas via the explicit drag term, introducing another transport term in the direction of gas density gradients (see Eq.~\ref{eq:mean_mass_flux_Reynolds}). The combination of these two effects results in directed transport against gradients of dust-to-gas ratio, as predicted by Eq.~\ref{eq:steady_state_transport_flux}. Note, in a uniform background ($\rho_g=\mathrm{const.}$, $t_t=\mathrm{const.}$), Eq.~\ref{eq:steady_state_transport_flux} simplifies to the classical gradient diffusion flux of Eq.~\ref{eq:GDH}. \newline
Interestingly, in a non-uniform background, there is a novel transport term arising from gradients of $t_t$ in Eq.~\ref{eq:steady_state_transport_flux}, that is not predicted by any of the gradient diffusion transport models. Therefore, we now aim to confirm this prediction by means of a direct comparison to the Lagrangian turbulence model introduced in Sec.~\ref{sec:Stochastic Lagrangian Formalism}. Specifically, we perform a numerical experiment and study the turbulent spreading of a population of dust grains in a nonuniform, but static ($\bar{u}_i=\bar{v}_i=0$) gas background.\newline 
We set up a numerical experiment such that the gas density in the background is constant, to eliminate possible contributions from the first term on the r.h.s. of Eq.~\ref{eq:steady_state_transport_flux}. However, we still allow the stopping time $t_s$ to vary in space, e.g., via variations in the speed of sound. Note, we do not discuss the possible physical feasibility of such a setup here, since our experiment is purely numerical in nature.\newline
In our first fiducial example, we set the correlation time and the stopping time to constant values ($t_\mathrm{corr}=0.01$ and $t_s=t_{s,0}=1$ in arbitrary units) such that the stopping time $t_s$ is long compared to the correlation time $t_\mathrm{corr}$ and consequently $t_t\simeq t_s$, and the background is modeled to be uniform. Further, we set the diffusion coefficient to a constant value of $D=10^{-3}$ and numerically solve the stochastic equations of motion (Eqs.~\ref{eq:lagr_1}, \ref{eq:lagr_3}, \ref{eq:SEOM2}, and \ref{eq:SEOM3}) for a number of $N=2\cdot10^3$ particles, initially at rest at $x=0$, with an explicit Euler scheme and a numerical timestep $\Delta t = 0.01\cdot t_\mathrm{corr}$. \newline
Because the stopping time $t_s$ is constant in this first setup, we do not expect a systematic drift, only symmetric diffusive spreading of the particle population. We confirm this by showing the temporal evolution of the particles in the reference setup in the upper subplot of Fig. \ref{fig:random_walk}. In the figure, gray background colors show, for each time $t$, the normalized kernel density estimate of the particle positions in $x$-$t$-space. The solid red line shows the mean of the distribution $\langle x \rangle$ at each point in time, and the red-shaded region covers the region within one standard deviation of the mean value. Over time, the distribution diffusively spreads but remains centered around $x=0$. This is the expected effect of random turbulent motions of particles in a uniform background \citep{Visser97}.\newline
In our second example, we allow the stopping time $t_s$ to vary depending on the particle position, to model the motion of particles through a nonuniform gas background. We choose the stopping time to exponentially increase towards increasing values of $x$ as $t_s=t_{s,0}\exp(3x)$, such that if the solution indeed follows Eq.~\ref{eq:turbulent_mass_flux_Favre}, the expected systematic drift velocity is independent of $x$. \newline
We show the solution with the varying stopping time in the lower subplot of Fig. \ref{fig:random_walk} where we plot the mean particle position with a solid blue line and shade the region within one standard deviation in blue color. We find the width of the particle distribution to spread diffusively, as in the example with a uniform background, but the mean of the distribution $\langle x \rangle$ drift towards increasing values of the turbulent timescale $t_t\simeq t_s$ with a systematic and constant velocity $v=D\nabla \ln t_s$. The mean of the particle distribution coincides with the green dashed line, which represents the prediction of Eq.~\ref{eq:turbulent_mass_flux_Favre}. \newline
We conclude that, in a nonuniform gas background with isotropic and homogeneous turbulence ($D=\mathrm{const.}$, $t_\mathrm{corr}=\mathrm{const.}$), and for large particles such that $t_t\simeq t_s$, there exists systematic transport of particles towards increasing values of the stopping time $t_s$ as described by Eq.~\ref{eq:turbulent_mass_flux_Favre}. In other words, in this regime, the turbulent flux is non-Fickian, an effect which is not captured by classical gradient diffusion models. \newline
In the following section, we focus on applications of the model to protoplanetary disks. 

\subsection{The Vertical Steady-State Disk Profile Revisited}
\label{sec:Vertical Steady-State Disk Profile}
\begin{figure}
\centering
\includegraphics[width=1.0\columnwidth]{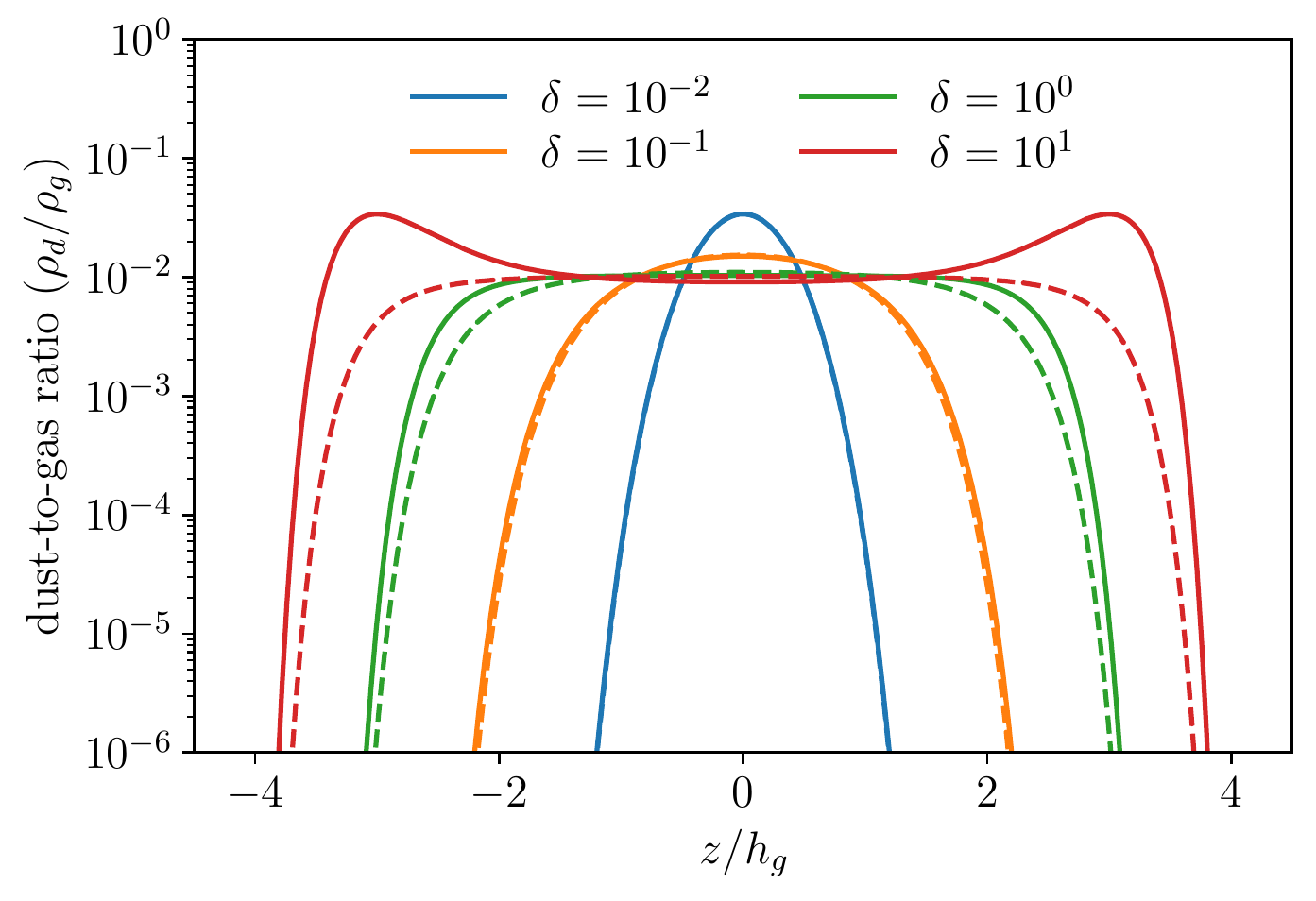}
\caption[Vertical steady-state profile of the dust-to-gas ratio.]{Vertical steady-state profile of the dust-to-gas ratio. The solid lines represent Eq.~\ref{eq:hom_solution} for $St_\mathrm{mid}=0.1$ and $t_\mathrm{corr}\Omega=1$, and different values of the diffusivity $\delta$. The vertical gas profile is assumed Gaussian and the surface density ratio is 1:100. The dashed lines follow the gradient diffusion solution of Eq.~\ref{eq:FromangandNelson}. The different values of the diffusivity correspond to $u'^2/c_s^2=0.01,0.1,1,10$ where $u'^2=D/t_\mathrm{corr}$ represents the squared turbulent velocity dispersion. The two solutions differ only when the turbulent velocity dispersion approaches the speed of sound $u'^2 \gtrsim c_s^2$, in regions where dust grains decouple from turbulent eddies. For subsonic turbulence $u'^2 \ll c_s^2$, the two solutions are indistinguishable.}
\label{fig:vertical_profile}
\end{figure}

In this section, we apply the turbulent particle pressure model to find an analytical solution for the vertical equilibrium profile of a protoplanetary disk. For this, we solve the system of equations along dimension $z$ under the influence of the vertical component of the stellar gravitational field $g_z=-\Omega^2z$. For the system to be static, the time derivatives and also the velocities vanish $\tilde{v}_z=\tilde{u}_z=0$. Note that a static solution only requires the Favre-averaged velocities to vanish, such that the net flux is zero ($\bar{\rho}_d \tilde{v}_z=0$) and the mean and turbulent fluxes cancel each other out ($\bar{\rho}_d \bar{v}_z+\bar{\rho}_d v^*_z=0$), but it does not necessarily require $\bar{v}_z$ or $v^*_z$ to vanish. \newline
The particle mass conservation equation is then fulfilled trivially, and the system of particle equations that we must solve is
\begin{subequations}
\begin{equation}\label{eq:vert_equi_a} 
    0=-\frac{1}{t_s}\bar{\rho}_d\bar{v}_z+\frac{D}{t_s}\frac{\bar{\rho}_d}{\bar{\rho}_g} \frac{\partial}{\partial z} \bar{\rho}_g-\bar{\rho}_d\Omega^2z
\end{equation}
\begin{equation}\label{eq:vert_equi_b} 
    \frac{\partial}{\partial z}\bigg(\frac{D}{t_t}\bar{\rho}_d\bigg)=-\frac{1}{t_t}\bar{\rho}_dv^*_z
\end{equation}
\end{subequations}
From $\tilde{v}_z=0$ and Eq.~\ref{eq:Rey_Favre_connection}, it follows that $\bar{v}_z=-v^*_z$. Without yet making use of the fact that $t_\mathrm{corr}$ and $D$ are constant, we reduce the system of Eq.~\ref{eq:vert_equi_a} and Eq.~\ref{eq:vert_equi_b} to the following partial differential equation, which describes the vertical dust equilibrium profile of a protoplanetary disk:
\begin{equation}\label{eq:pde_vertical_disk_profile}
    \frac{\partial}{\partial z}\bigg[\ln\bigg(\frac{D}{t_t}\frac{\rho_d}{\rho_g}\bigg)\bigg]=-\frac{\Omega^2 t_s}{D}z
\end{equation}
Assuming that the vertical profile of the gas background is Gaussian with scale height $h_g$, the solution to Eq.~\ref{eq:pde_vertical_disk_profile} becomes 
\begin{equation}\label{eq:hom_solution}
\begin{split}
    \rho_d(z)=&\rho_{d,0}\:\bigg[1+\frac{t_{s,\mathrm{mid}}}{t_\mathrm{corr}}\exp\bigg({\frac{z^2}{2h_g^2}}\bigg)\bigg] \times \\
    & \exp\bigg[ -\frac{\Omega t_{s,\mathrm{mid}}}{\delta}\bigg( \exp\bigg(\frac{z^2}{2h_g^2} \bigg)-1\bigg)-\frac{z^2}{2h_g^2}\bigg]
\end{split}
\end{equation}
where $t_{s,\mathrm{mid}}$ is the stopping time evaluated at the disk midplane. \newline
For a turbulent velocity dispersion, which is comparable to the sound speed ($u'^2 = D/t_\mathrm{corr}\gtrsim c_s^2$), we find the novel factor $1+t_s/t_\mathrm{corr}$ in Eq.~\ref{eq:hom_solution} to locally increase the dust density in regions where the stopping time is comparable or larger than the correlation time. This increase is due to the transition from drag dominated to inertia-dominated dust dynamics in the disk atmosphere. \newline
{For subsonic turbulence ($D/t_\mathrm{corr}\ll c_s^2$), we find the difference between Eq.~\ref{eq:hom_solution} and Eq.~\ref{eq:FromangandNelson} to be vanishingly small. We illustrate this in Fig.~\ref{fig:vertical_profile}, where we plot the vertical profile of the dust-to-gas ratio calculated with Eq.~\ref{eq:hom_solution} assuming a Gaussian gas profile with scale height $h_g^2 = (c_s^2+D/t_\mathrm{corr})/\Omega^2$ in solid lines for different values of the diffusivity $\delta$ and $t_\mathrm{corr}\Omega=1$, and $St_\mathrm{mid}=0.1$. The different values of the diffusivity correspond to a turbulent velocity dispersion ranging from subsonic to supersonic turbulence ($u'^2/c_s^2=0.01,0.1,1,10$). For comparison, we also calculate the profile with Eq.~\ref{eq:FromangandNelson} in dashed lines. \newline
as predicted, the two solutions differ only when the turbulent velocity dispersion approaches the speed of sound $u'^2 \gtrsim c_s^2$, and in regions where dust grains decouple from turbulent eddies.} For subsonic turbulence $u'^2 \ll c_s^2$, which is expected in protoplanetary disks, the two solutions are indistinguishable. \newline
Focusing on subsonic turbulence, in the limit of small particles $(St \to 0)$, the vertical scale height of the dust $h_d$ approaches the scale height of the gas ($h_d \to h_g$) and thus fulfills the good mixing condition. \newline
Close to the disk midplane ($z\ll h_g$), the vertical static equilibrium profile (Eq.~\ref{eq:hom_solution}) is approximately Gaussian. We Taylor expand Eq.~\ref{eq:hom_solution} up to the second order in $z$, and write for small values of $z$ ($z\ll h_g$) the ratio of the scale heights as: 
\begin{equation}\label{eq:ex_sclae_height_ratio}
   \frac{h_d^2}{h_g^2}\approx \frac{\delta}{\delta+St}
\end{equation}
where we have assumed subsonic turbulence $D/t_\mathrm{corr}\ll c_s^2$, which is equivalent to the relation $\delta \ll \Omega t_\mathrm{corr}$. We highlight that the correlation time $t_\mathrm{corr}$ does not appear in Eq.~\ref{eq:ex_sclae_height_ratio} as a result of assuming subsonic turbulence. \newline
To our knowledge, we present for the first time, a self-consistent derivation of the vertical settling-diffusion equilibrium profile that correctly captures the small and large particle limits, without the use of a heuristic argument. \newline
Next, we derive an effective vertical diffusion coefficient $D_{d,z}^\mathrm{eff}$ analogous to \citet{Carballido11}. We assume the dust scale height to be small compared to the gas scale height ($h_d\ll h_g$), such that the dust settles into a thin region close to the midplane in which the gas density is basically constant in the vertical direction. From Eq.~\ref{eq:mean_gas_velc}, it then follows that $\bar{u}_z=0$ and, we can safely neglect the interactions of the dust with the gas via the explicit drag term that would be present in a nonuniform gas background. From the condition $h_d\ll h_g$ and Eq.~\ref{eq:ex_sclae_height_ratio}, it follows that
\begin{equation}\label{eq:large_frain_limit_shr}
    \frac{h_d^2}{h_g^2}=\frac{\delta}{St}
\end{equation}
We then set the diffusion timescale $t_\mathrm{diff}$ (Eq.~\ref{eq:diffusion timescale}) across the dust scale height, i.e. $t_\mathrm{diff}=h_d^2/D_{d,z}^\mathrm{eff}$, equal to the vertical settling time $t_\mathrm{sett}$. To estimate the settling time, we note that large particles ($St \gg 1$) undergo damped vertical oscillation with a settling time $t_\mathrm{sett}=St/\Omega$. Tightly coupled particles ($St \ll 1$) obtain terminal velocity and settle in a time  $t_\mathrm{sett}= 1/\Omega St$. Combining these two results gives \citep{Youdin2007}:
\begin{equation}\label{eq:settling_time_comp}
    t_\mathrm{sett}\approx \frac{St}{\Omega}+\frac{1}{\Omega St}
\end{equation}
Combining Eq.~\ref{eq:large_frain_limit_shr} and Eq.~\ref{eq:settling_time_comp} to calculate an effective diffusion coefficient gives 
\begin{equation}\label{eq:eff_d_d_z}
    D_{d,z}^\mathrm{eff} = \frac{D}{1+St^2}
\end{equation}
in agreement with \citet{Carballido11} and \citet{Youdin2007}.\newline

\subsection{Large Grains Limit}
\label{sec:Large Grains Limit}
In the limit of large grains ($t_s \gg t_\mathrm{corr}$), the two momentum equations, Eq.~\ref{eq:momentum_like_aa} and Eq.~\ref{eq:momentum_like_bb}, can be combined to one equation which reads
\begin{equation}\label{eq:small_t_corr_limit} 
\begin{split}
    \frac{\partial}{\partial t}\big(\bar{\rho}_d\tilde{v}_i\big)+\frac{\partial}{\partial x_j}\bigg( \bar{\rho}_d\tilde{v}_i \tilde{v}_j+\frac{1}{3}\frac{D}{t_s}\bar{\rho}_d\delta_{ij}\bigg)=&
    -\frac{1}{t_s}\bar{\rho}_d\big(\tilde{v}_i-\bar{u}_i\big)+\\
    &\frac{D}{t_s}\frac{\bar{\rho}_d}{\bar{\rho}_g} \frac{\partial}{\partial x_i} \bar{\rho}_g.
\end{split}
\end{equation}
In general, Eq.~\ref{eq:momentum_like_aa} and Eq.~\ref{eq:momentum_like_bb} can always be combined to give Eq.~\ref{eq:Favre_mom_eq}, but only in this limit of large grains, the two velocities $\bar{v}_i$ and $v^*_i$ can be eliminated and Eq.~\ref{eq:small_t_corr_limit} can be written in terms of  $\tilde{v}_i$ only. Consequently, a second dust momentum equation is not needed anymore for the system of Eq.~\ref{eq:favre_mass_eq} and Eq.~\ref{eq:small_t_corr_limit} to be closed, and the system of equations is simplified. \newline
In a uniform gas background ($\rho_g=\mathrm{const.}$), the second term on the r.h.s of Eq.~\ref{eq:small_t_corr_limit} vanishes and the equation is formally identical to Eq.~\ref{eq:Klahr_mom_cons}. Therefore, the model of \citet{Klahr2021}, can be interpreted as the large-grain limit of our more general model.

\section{Linear Perturbation Analysis}
\label{sec:LPA}
We perform a linear perturbation analysis analogous to the analysis in \citet{Binkert23b} to investigate the linear dynamics of our novel turbulent transport model and to identify key characteristics of the turbulent transport model derived in this work. We first study a one-dimensional problem in Sec.~\ref{sec:One Dimension Without External Forces}, before we focus on a two-dimensional and axisymmetric disk in Sec.~\ref{sec:Axisymmetric Keplerian Disks}.

\begin{figure}
\centering
\includegraphics[width=1.0\columnwidth]{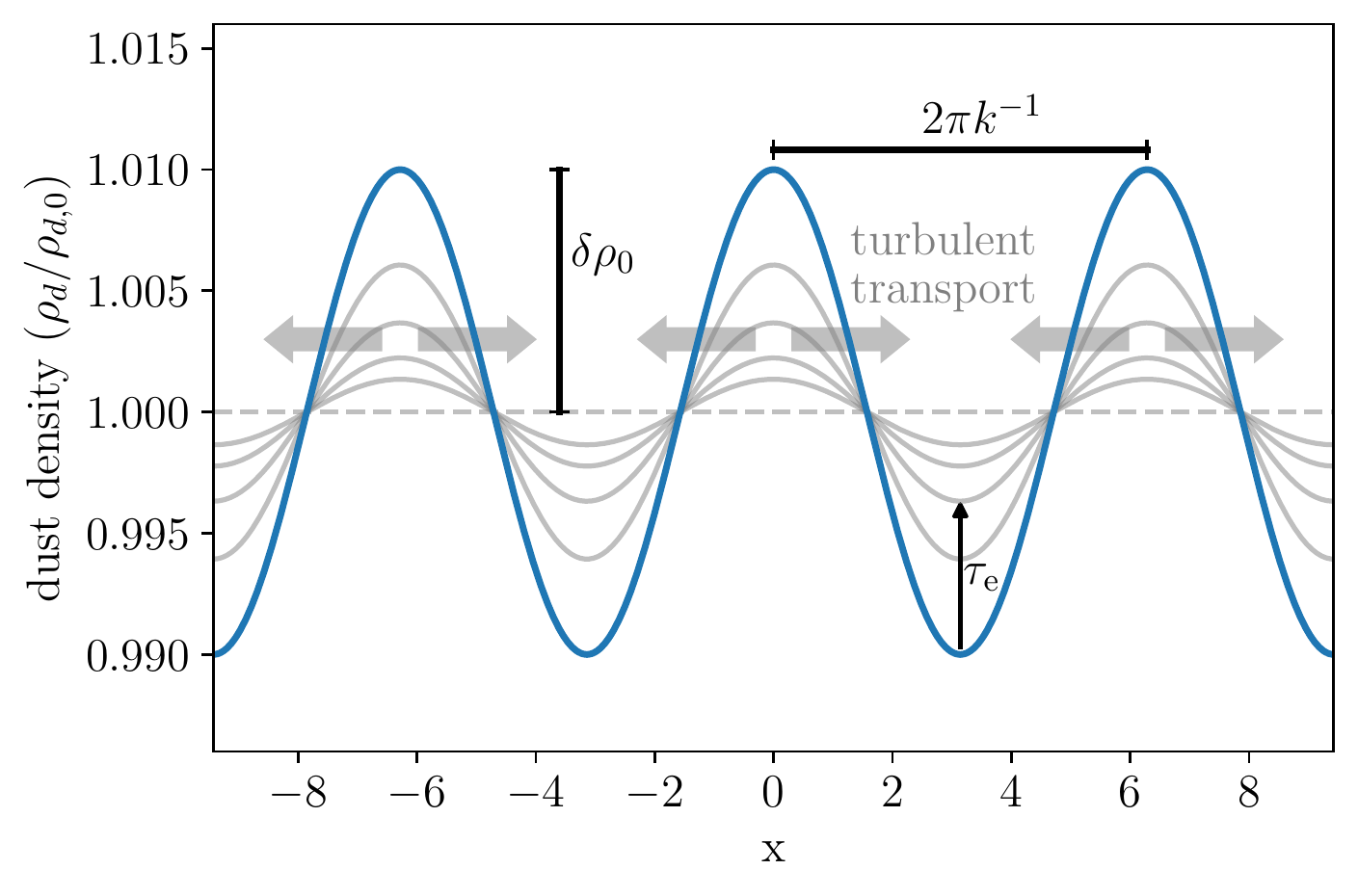}
\caption[This figure illustrates the decay of a harmonic perturbation.]{This figure illustrates the turbulent decay of a harmonic perturbation to the dust density $\rho_d$, as analyzed in the linear perturbation analysis in Sec. \ref{sec:LPA}, in arbitrary units. The perturbation is characterized by its amplitude $\delta\rho_0$, which is small compared to the background, $\delta\rho_0 \ll \rho_{d,0}$. Its wavenumber $k$ is related to the wavelength of the perturbation as $\lambda = 2\pi k^{-1}$. The blue line represents the initial state of the perturbation, while the gray lines represent the decaying solutions at every half e-folding time, $\tau_e$. An effective diffusion coefficient $D^\mathrm{eff}$ can be calculated as the inverse of the product of the e-folding time $\tau_e$, and the square of the perturbation's wavenumber, $D^\mathrm{eff}=\tau_e^{-1} k^{-2}$. An effective diffusion coefficient can be calculated for any decaying perturbation, but only in a purely diffusive solution $D^\mathrm{eff}$ is independent of the wavenumber $k$.}
\label{fig:diffusion_sketch}
\end{figure}

\begin{figure*}
\centering
\includegraphics[width=1.8\columnwidth]{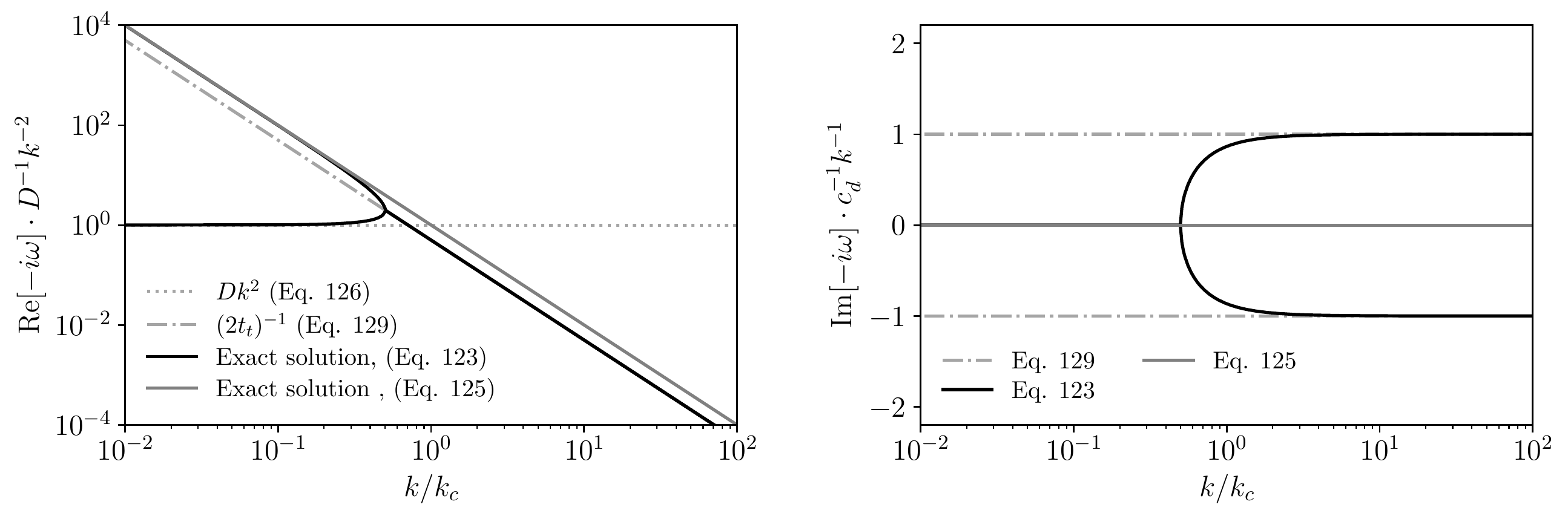}
\caption[Decay rates $-i\omega(k)$ of harmonic perturbations in one dimension in the absence of external forces.]{Decay rates $-i\omega(k)$ of harmonic perturbations in one dimension in the absence of external forces, i.e., the solution to the dispersion relation in Eq.~\ref{eq:disp_relation_system3}. Shown here is a case $t_\mathrm{corr} \ll t_s$ such that $t_t\approx t_s$. \textit{Left:} Shows the real part of the decay rates ($\mathrm{Re}[-i\omega(k)]$) normalized by the factor $D k^{2}$, such that diffusive solutions are represented by lines with slope zero. The black solid lines represent the two exact solutions of Eq.~\ref{eq:solution_disp_relation_system3}. The gray solid line represents the third exact solution in Eq.~\ref{eq:solution_disp_relation_system4}. For $k^2\gg k_c^2$, there exists no diffusive solution. The gray dotted line traces the diffusive solution (Eq.~\ref{eq:second_1d_eigenvalue_1}) that coincides with the exact solution on small scales ($k^2\ll k_c^2$). The gray dash-dotted line represents the solution in Eq.~\ref{eq:disp_relation_system3_2}. \textit{Right:} Imaginary part of the decay rate ($\mathrm{Im}[-i\omega]$) normalized by a factor $c_d k$. A non-zero value represents a traveling wave solution, and a zero-slope line represents solutions traveling at the same speed. The black solid lines show the exact solutions to Eq.~\ref{eq:disp_relation_system3}. The gray dot-dashed line follows the two analytic solutions of Eq.~\ref{eq:disp_relation_system3_2}, valid for $k^2\gg k_c^2$. For small wave numbers ($k^2\ll k_c^2$), these solutions are not travelling ($\mathrm{Im}[i\omega]=0$), as predicted by the explicit solutions in Eq.~\ref{eq:second_1d_eigenvalue} and Eq.~\ref{eq:third_1d_eigenvalue}.}
\label{fig:1d_disp_relation}
\end{figure*}

\subsection{One Dimension Without External Forces}
\label{sec:One Dimension Without External Forces}
We model dust in a turbulent and uniform gaseous background (with constant $\bar{\rho}_g$) in one dimension along the $x$-axis. We assume the gas to be static ($\tilde{u}=0$). From Eq.~\ref{eq:mean_gas_velc}, it then follows that $\bar{u}=0$ and consequently $u^*=0$. As always in this work, we assume the turbulence to be characterized by a constant diffusion coefficient $D$ and constant correlation time $t_\mathrm{corr}$. We describe the dust fluid using the linearized forms of Eq.~\ref{eq:favre_mass_eq} and Eq.~\ref{eq:Favre_mom_eq} and assume the dust-to-gas ratio to be small $\bar{\rho}_d/\bar{\rho}_g\ll 1$, such that the dust does not affect the gas. We will introduce small harmonic perturbations to the linearized dust density and velocity equations on top of a static background distribution. \newline
The set of linearized equations is as follows:
\begin{equation}
    \frac{\partial \bar{\rho}_d}{\partial t}+\bar{\rho}_d\frac{\partial }{\partial x}(\bar{v}+v^*)=0
\end{equation}
\begin{equation}
    \frac{\partial \bar{v}}{\partial t} = -\frac{\bar{v}}{t_s}-\frac{D}{t_t\bar{\rho}_d}\frac{\partial \bar{\rho}_d}{\partial x}
\end{equation}
\begin{equation}
    \frac{\partial v^*}{\partial t} = -\frac{v^*}{t_t}
\end{equation}
We describe the perturbed dust density as $\bar{\rho}_d=\bar{\rho}_{d,0}+\delta \rho_d$, where the perturbation $\delta \rho_d$ is small compared to the background $\bar{\rho}_{d,0}$. This perturbation to the dust density is illustrated in Fig. \ref{fig:diffusion_sketch}. We also introduce a perturbation to the turbulent transport velocity $v^*=\delta v^*$ and to the mean velocity $\bar{v}=\delta \bar{v}$. The perturbations are harmonic and have the form
\begin{equation}\label{eq:pert_rho_lin}
    \delta \rho_d = \delta \rho_{0}e^{i(\omega t+kx)}
\end{equation}
\begin{equation}
    \delta \bar{v} = \delta \bar{v}_0e^{i(\omega t+kx)}
\end{equation}
\begin{equation}\label{eq:pert_v_star_lin} 
    \delta v^* = \delta v^*_0e^{i(\omega t+kx)}
\end{equation}
where $\omega$ is the frequency and $k$ is the wavenumber of the perturbations.\newline
Plugging in the perturbed quantities into the linearized equations, and considering at most first-order terms, we find the following equations that we represent as a three-dimensional matrix equation as follows:
\begin{equation}\label{eq:1d_matrix_eq}
i\omega
    \begin{pmatrix}
 \delta \rho_d\\
\delta \bar{v}\\
\delta v^*
\end{pmatrix}
= 
    \begin{pmatrix}
0 & -ik\bar{\rho}_{d,0} & -ik\bar{\rho}_{d,0} \\
0 & -t_s^{-1} & 0\\
-ik\frac{D}{t_t \bar{\rho}_{d,0}} & 0 & -t_t^{-1}
\end{pmatrix}
    \begin{pmatrix}
  \delta \rho_d\\
\delta \bar{v}\\
\delta v^*
\end{pmatrix}
\end{equation}
The dispersion relation of the equation above reads
\begin{equation}
\label{eq:disp_relation_system3}
    \Big(i\omega\big(i\omega+t_t^{-1}\big)+D k^2t_t^{-1}\Big)(i\omega + t_s^{-1}) = 0
\end{equation}
which has two symmetric solutions
\begin{equation}\label{eq:solution_disp_relation_system3}
  i\omega_{1,2}=-\frac{1}{2t_t}\Bigg(1\pm \sqrt{1-4 \frac{k^2}{k_c^2}} \Bigg)
\end{equation}
where we have defined the \textit{characteristic wavenumber} $k_c$ as 
\begin{equation}\label{eq:character_wave_number}
    k_c^2\equiv D^{-1}t_t^{-1}
\end{equation}
and a third solution 
\begin{equation}\label{eq:solution_disp_relation_system4}
  i\omega_{3}=-t_s^{-1}
\end{equation}
A solution to the dispersion relation $i\omega$ is called the \textit{growth rate} of the perturbation. Conversely, $-i\omega$ is called the \textit{decay rate}. In Fig.~\ref{fig:1d_disp_relation}, we plot the decay rates $-i\omega(k)$, i.e., the solutions to Eq.~\ref{eq:disp_relation_system3}, as solid lines for the special case $t_\mathrm{corr} \ll t_s$.

\subsubsection{Dynamics on Large Scales (small wave numbers $k^2\ll k_c^2$)}
We analytically study the growth rates on large scales, i.e., small wave numbers $k^2\ll k_c^2$, where the three solutions to the dispersion relation in Eq.~\ref{eq:disp_relation_system3} are real-valued and negative and can be approximated by
\begin{equation}\label{eq:second_1d_eigenvalue_1}
  i\omega_1=-Dk^2
\end{equation}
\begin{equation}\label{eq:second_1d_eigenvalue}
  i\omega_2=-t_t^{-1}
\end{equation}
\begin{equation}\label{eq:third_1d_eigenvalue}
  i\omega_3=-t_s^{-1}
\end{equation}
The three growth rates represent decaying solutions because they are all real-valued and negative. We plot these approximate solutions as gray discontinuous lines as a function of the wave number $k$ in Fig.~\ref{fig:1d_disp_relation} on top of the exact solutions.

\subsubsection{Dynamics on Small Scales (large wave numbers $k^2\gg k_c^2$)}
On small spatial scales, i.e., for large wave numbers, the first two solutions to the dispersion relation in Eq.~\ref{eq:disp_relation_system3} can be approximated by 
\begin{equation}\label{eq:disp_relation_system3_2}
  i\omega_{1,2}=-\frac{1}{2t_t}\pm i k c_d
\end{equation} 
We plot the above growth rate as a gray dot-dashed line as a function of the wave number $k$ in Fig.~\ref{fig:1d_disp_relation}. It coincides with the exact solution (black line) for $k^2\gg k_c^2$. The imaginary part of Eq.~\ref{eq:disp_relation_system3_2} indicates that the eigensolutions are traveling waves that propagate at speed $c_d=\sqrt{D/t_t}$. The speed of propagation is identical to the value of the turbulent particle dispersion (Eq.~\ref{eq:turb_vel_disp_def}). This traveling wave solution was first discussed in \citet{Klahr2021}, but their model predicts a non-physical supersonic wave speed for short stopping times $t_s<D/c_s^2$ and thus must be modified in their model. In this model, the propagation speed $c_d$ does never exceed the turbulent gas velocities, even for short stopping times (see Eq.~\ref{eq:turb_vel_disp_def}). 
\newline
We write the real parts of the eigensolutions that correspond to the first two eigenvalues in Eq.~\ref{eq:disp_relation_system3_2} explicitly as 
\begin{equation}\label{eq:TPPM_eigensol_b}
    \mathrm{Re}(\delta \rho_d) =\pm\delta v^*_0\frac{\bar{\rho}_{d,0}}{D}\cos(kx \pm k c_d t)e^{-t/2 t_t}
\end{equation}
\begin{equation}\label{eq:TPPM_eigensol_a}
   \mathrm{Re}(\delta v^*) = \delta v^*_0\cos(kx \pm k c_d t )e^{-t/2 t_t}
\end{equation}
\begin{equation}\label{eq:TPPM_eigensol_c}
   \mathrm{Re}(\delta \bar{v}) = 0
\end{equation}
These expressions indeed describe waves traveling at speed $c_d$ and decaying on a timescale $\tau=2 t_t$. The third solution to the dispersion relation in Eq.~\ref{eq:disp_relation_system3} has the same form on small scales as on large scales, and decays on a timescale equal to the stopping time without an oscillating imaginary component:
\begin{equation}\label{eq:solution_disp_relation_system42}
  i\omega_{3}=-t_s^{-1}
\end{equation}

\subsubsection{Physical Interpretation of Turbulent Dust Transport in 1D}
\label{sec:Physical Interpretation of Turbulent Dust Transport in 1D}
On large scales ($k^2\ll k_c^2$), the first eigenvalue in Eq.~\ref{eq:second_1d_eigenvalue_1} represents a non-oscillating perturbation that decays on an e-folding time $\tau_{1}=D^{-1}k^{-2}$, which is equivalent to a diffusive decay characterized by diffusion coefficient $D$. The corresponding eigensolutions ($\delta v^*$, $\delta {\rho}_\mathrm{d}$) to this eigenvalue fulfill the relation
\begin{equation}
     \bar{\rho}_\mathrm{d,0} \delta v^* = -ikD\delta \rho_\mathrm{d}
\end{equation}
Using Eq.~\ref{eq:pert_rho_lin}, this expression can be rewritten as 
\begin{equation}\label{eq:linear_force balance}
     \bar{\rho}_\mathrm{d} \delta v^* = -D\frac{\partial}{\partial x} \bar{\rho}_\mathrm{d}
\end{equation}
It is apparent that Eq.~\ref{eq:linear_force balance} represents the gradient diffusion equilibrium flux under force balance (compare to Eq.~\ref{eq:turbulent_mass_flux_Favre}) and thus confirms the diffusive nature of this eigensolution. \newline
The second and third eigensolutions corresponding to the eigenvalues in Eq.~\ref{eq:second_1d_eigenvalue} and in Eq.~\ref{eq:third_1d_eigenvalue}, do not fulfill the force balance given by Eq.~\ref{eq:turbulent_mass_flux_Favre}. Instead, the eigensolutions represent perturbations that evolve towards restoring the force balance on an e-folding time $\tau_{2}=t_t$ and $\tau_{3}=t_s$, respectively. \newline 
To summarize, on large scales ($k^2\ll k_c^2$), we have found three characteristic solutions. The first solution represents a diffusive decay of a perturbation under force balance on a timescale $\tau_{1}=D^{-1}k^{-2}$. The other two solutions represent the decay of an out-of-equilibrium perturbation.\newline
On large scales ($k^2\ll k_c^2$), the decay of the out-of-equilibrium perturbation is much faster than the diffusive decay of the first solution because, according to Eq.~\ref{eq:character_wave_number}, on large scales, $t_t \ll D^{-1}k^{-2}$, and also $t_s \ll D^{-1}k^{-2}$, hold. This indicates that the long-time evolution of large-scale perturbations is dominated by diffusive processes.\newline
On small scales ($k^2\gg k_c^2$), the decay rates either follow the decaying wave solution of Eq.~\ref{eq:disp_relation_system3_2} or the decaying non-traveling solution of Eq.~\ref{eq:solution_disp_relation_system42}, which are both independent of the wavenumber $k$, indicating that on small scales, there exists no diffusive solution. \newline
This property of our turbulent pressure model is distinctly different from a gradient diffusion model, in which diffusive solutions, per definition, exist on all scales \citep[see e.g. appendix A of ][]{Binkert23b}. \newline
From a physics point of view, a diffusive solution does not exist on small scales because momentum cannot be transferred from the turbulent gas to the dust on timescales smaller than $t_t$ (or vice versa). A purely diffusive solution would require perturbation to decay on a timescale $\tau=D^{-1}k^{-2}$ which is per definition smaller than $t_t$ for wave numbers larger than $k_c$.\newline
We conclude that small-scale perturbations (for wave numbers $k^2\gg k_c^2$) in this model survive for longer than a purely diffusive evolution would predict, due to the finite coupling of the dust to turbulence. \newline
\citet{Binkert23b} illustrate the difference between the gradient diffusion model and the pressure-driven turbulent transport model (in the case $t_\mathrm{corr}\ll t_s$) in their Figure 1 in an example of a decaying Gaussian perturbation.\newline
{Lastly, we aim to provide an intuitive explanation for the traveling wave solution, which is somewhat unexpected to occur in the originally pressureless dust fluid. This behavior mirrors a sound wave in gas because particles in a high-density region spread out towards an equilibrium distribution, driven by turbulent mixing, similar to how gas molecules in a high-pressure region spread due to thermal pressure. In our model, turbulent dust fluxes carry momentum and thus have inertia. The inertia of the dust particles causes the particles to overshoot their equilibrium distribution, creating another overdensity. This process then restarts and can be described as a wave. Ultimately, the wave solution is a result of turbulence-driven rarefaction and consequent compression from inertia. Further investigations should confirm if these traveling waves are physical or just artifacts of the Reynolds/Favre averaging process.}

\begin{figure*}
\centering
\includegraphics[width=1.8\columnwidth]{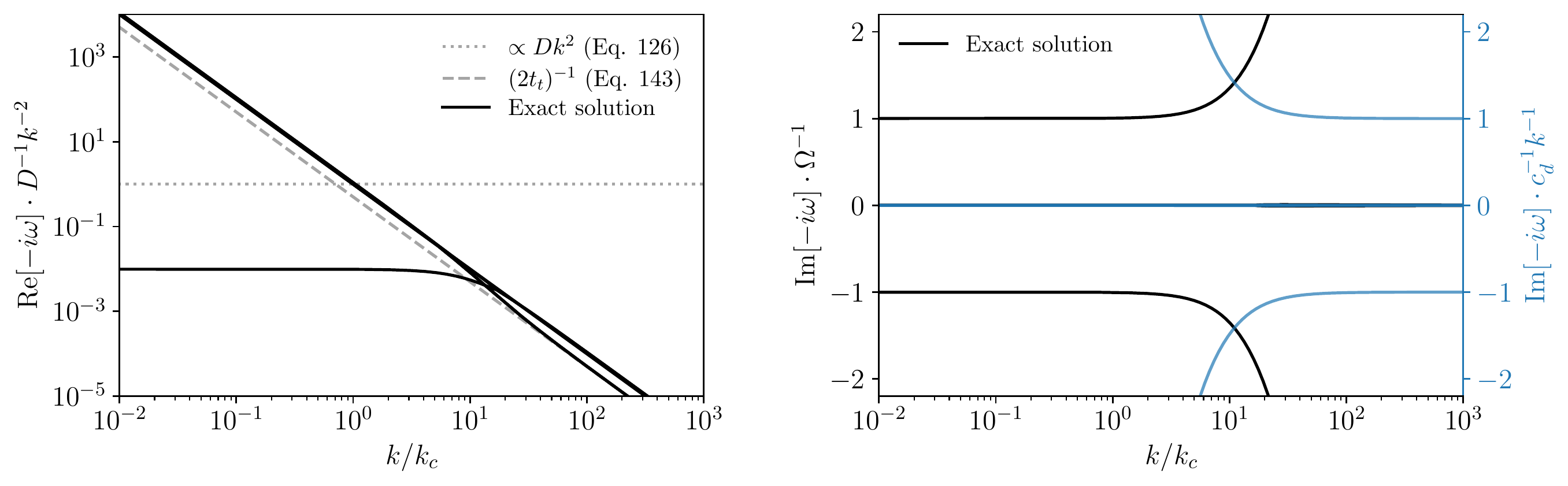}
\caption[Decay rates $-i\omega(k)$ of harmonic perturbation in a two-dimensional, axisymmetric Keplerian disk.]{We illustrate the decay rates $-i\omega(k)$ of harmonic perturbation to the dust density in a two-dimensional, axisymmetric Keplerian disk. Shown is the numerical solution to the dispersion relation of Eq.~\ref{eq:2d_Matrix} for $\Omega t_\mathrm{corr}=1$ and $St = 10$. \textit{Left:} Shows the real part of the decay rates ($\mathrm{Re}[-i\omega]$), normalized by the factor $D k^{2}$, such that diffusive solutions are represented by lines with slope zero. The black solid lines represent the exact solutions. The gray dashed lines represent the solution given by $-i\omega = (2t_t)^{-1}$ as in Eq.~\ref{eq:small_scale_sol_2}. The gray dotted line represents a diffusive solution $-i\omega = Dk^2$. Note, how the actual diffusive solution (horizontal black line) decays a factor 100 slower than expected due to the effects of epicyclic oscillations and the factor $1/(1+t_t^2\Omega^2)$ reducing the effective diffusivity in radial direction (Eq.~\ref{eq:small_scale_sol_5}). \textit{Right:} Imaginary part of the normalized decay rate ($\mathrm{Im}[-i\omega]$). The black solid lines represent the exact solutions and are normalized by a factor $\Omega^{-1}$ such that the epicyclic frequency represents a horizontal line at $\mathrm{Im}[-i\omega]\Omega^{-1}=\pm 1$. The blue solid line represents the same solution but is normalized by a factor $c_d k$ such that a horizontal line represents wave solutions traveling at speed $c_d$.}
\label{fig:2d_disp_relation}
\end{figure*}

\subsection{Axisymmetric Keplerian Disk}
\label{sec:Axisymmetric Keplerian Disks}
After considering the one-dimensional case in the absence of external forces, we now consider a two-dimensional and axisymmetric Keplerian disk in the presence of gravity. We follow a dust fluid parcel along its orbit and use the linearized local shearing box approximation to describe its dynamics \citep{Goldreich65,Youdin11}. For this, we integrate the dynamical equations, i.e., Eq.~\ref{eq:mass_eq_22}, Eq.~\ref{eq:momentum_like_aa}, and Eq.~\ref{eq:momentum_like_bb}, along the vertical axis and rewrite the dynamical equations in local variables $r=r_0(1+x)$ such that $x\ll 1$, and in terms of the dust surface density $\bar{\Sigma}_d$ and the two velocity components $\bar{v}_r$, $v^*_r$ and $\bar{v}_\phi$, $v^*_\phi$. The five linearized equations of the axisymmetric system are as follows:
\begin{equation}
    \frac{\partial \bar{\Sigma}_d}{\partial t}+\bar{\Sigma}_d\frac{\partial}{\partial x} (v^*_r+\bar{v}_r)=0
\end{equation}
\begin{equation}
    \frac{\partial \bar{v}_r}{\partial t}-2\Omega (v^*_\phi+\bar{v}_\phi) = -\frac{\bar{v}_r}{t_s}
\end{equation}
\begin{equation}
    \frac{\partial \bar{v}_\phi}{\partial t}+\frac{1}{2} \Omega (v^*_r+\bar{v}_r)=-\frac{\bar{v}_\phi}{t_s}
\end{equation}

\begin{equation}
    \frac{\partial v_r^*}{\partial t}= -\frac{v^*_r}{t_t}-\frac{D}{t_t\bar{\Sigma}_d}\frac{\partial}{\partial x}\bar{\Sigma}_d
\end{equation}

\begin{equation}
    \frac{\partial v^*_\phi}{\partial t}=-\frac{v^*_\phi}{t_t}
\end{equation}
We introduce small perturbations, analogous to Eqs.~\ref{eq:pert_rho_lin}-\ref{eq:pert_v_star_lin}, in the radial direction to the dust surface density $\bar{\Sigma}_d=\bar{\Sigma}_{d,o}+\delta\Sigma_d$ and the radial and azimuthal components of the velocities ($\bar{v}_r=\delta \bar{v}_r$, $v_r^*=\delta v^*_r$, $\bar{v}_\phi=\bar{v}_{\phi,0}+\delta v_\phi$, $v_\phi^*=\delta v_\phi$). The azimuthal component of the mean velocity describes the Keplerian shear $\bar{v}_{\phi,0} = \Omega r_0(1-\frac{3}{2}x)$.\newline
We plug in the harmonic perturbations to the linearized equations and consider only first-order terms. The system in matrix notation reads
\begin{equation}\label{eq:2d_Matrix}
    i\omega    
    \begin{pmatrix}
 \delta \bar{\Sigma}_d\\
 \delta \bar{v}_r\\
  \delta \bar{v}_\phi\\
\delta v^*_r \\
\delta v^*_\phi
\end{pmatrix}
=
  \begin{pmatrix}
0 & -ik\bar{\Sigma}_{d,0} & 0 & -ik\bar{\Sigma}_{d,0} & 0\\
0 & -t_s^{-1} & 2\Omega & 0 & 2\Omega\\
0 & -\Omega / 2 & -t_s^{-1} & -\Omega/2 & 0\\
-\frac{ikD}{t_t \bar{\Sigma}_d} & 0 & 0 & -t_t^{-1} & 0\\
0 & 0 & 0 & 0 & -t_t^{-1}
\end{pmatrix}
\begin{pmatrix}
 \delta \bar{\Sigma}_d\\
 \delta \bar{v}_r\\
  \delta \bar{v}_\phi\\
\delta v^*_r \\
\delta v^*_\phi
\end{pmatrix}
\end{equation}
The fifth-order dispersion relation in Eq.~\ref{eq:2d_Matrix} is too complex to study analytically. Therefore, we determine its solutions numerically. In Fig.~\ref{fig:2d_disp_relation}, we plot the decay rates of the eigensolutions to the system in Eq.~\ref{eq:2d_Matrix} as black solid lines for values of $St=10$ and $t_\mathrm{corr}\Omega = 1$. \newline
Next, we study the dispersion relation in two limiting cases: large scales (in Sec.~\ref{sec:small_scales}) and small scales (in Sec.~\ref{sec:large_scales}). 

\subsubsection{Dynamics on Small Scales (large wave numbers)}
\label{sec:small_scales}
On small scales, i.e., for large wave numbers $k^2\gg k_c^2(1+St^2)$, we identify four traveling solutions with a constant decay rate (see Sec.~\ref{sec:small_scales} for an explanation of why there is an additional factor $1+St^2$). The first two solutions decay on a timescale equal to the stopping time $t_s$:
\begin{equation}\label{eq:small_scale_sol_1}
    \mathrm{Re}[i\omega_{1,2}]=-t_s^{-1}
\end{equation}
The third and the fourth solutions decay on a timescale equal to $2t_t$:
\begin{equation}\label{eq:small_scale_sol_2}
    \mathrm{Re}[i\omega_{3,4}]=-(2t_t)^{-1}
\end{equation}
The fifth and final solution is a non-traveling solution and decays on a timescale equal to $t_t$:
\begin{equation}\label{eq:small_scale_sol_3}
    i\omega_{5}=-t_t^{-1}
\end{equation}
For comparison with the exact solution across all scales, we illustrate the decay rates of the limiting solutions, as discussed here, on the l.h.s of Fig.~\ref{fig:2d_disp_relation}. For $t_s \gg t_\mathrm{corr}$, Eq.~\ref{eq:small_scale_sol_1} and Eq.~\ref{eq:small_scale_sol_2} are indistinguishable. Therefore, only two parallel lines appear in the lower right corner of the left subplot in Fig.~\ref{fig:2d_disp_relation}.\newline
Notably, there exists no diffusive solution on small scales ($k^2 \gg k_c^2(1+St^2)$) akin to the behavior in one dimension as discussed in Sec.~\ref{sec:One Dimension Without External Forces}. Assuming $t_t = \Omega^{-1}$, a disk aspect ratio of $h_g/r= 0.05$ and a diffusivity $\delta = 10^{-3}$, as appropriate for turbulent protoplanetary disks, the wave number $k_c$ corresponds to a length scale $2 \pi k_c^{-1}\sim0.2 h_g$, i.e., 20 per cent of the gas scale height. This is comparable to the vertical scale height of a dust disk with $St_\mathrm{mid} = 0.025$ ($h_d/h_g \sim \sqrt{\delta / St}\sim 0.2$).  

\subsubsection{Dynamics on Large Scales (small wave numbers)}
\label{sec:large_scales}
\begin{figure}
\centering
\includegraphics[width=0.99\columnwidth]{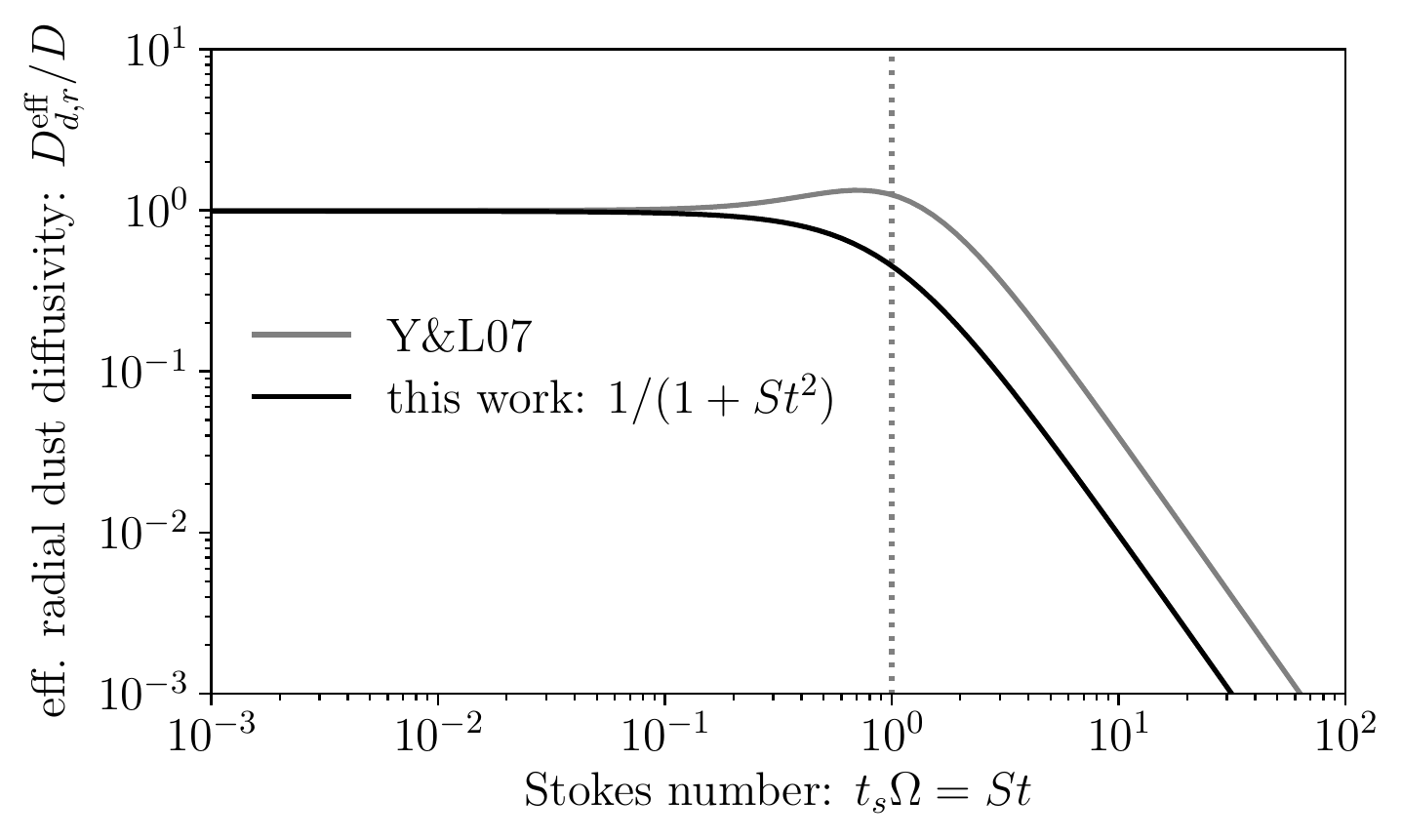}
\caption[Effective radial dust diffusion coefficient as a function of Stokes.]{Effective radial diffusion coefficient $D_{d,r}^\mathrm{eff}$ as a function of Stokes number in a two-dimensional Keplerian disk (as described by Eq.~\ref{eq:radial_dust_diff}). We set $k^2/k_c^2 =0.01$. The vertical dotted gray line represents the transition at $St\sim1$ below which the effective diffusion is constant. The solid gray line represents the solution of \citet{Youdin2007} (their Eq.~37). Ignoring corrections of order unity, our results describing turbulent transport in radial direction are consistent with the results of \citet{Youdin2007}.}
\label{fig:eff_diffusivity}
\end{figure}

On large scales, i.e., for small wave numbers ($k^2\ll k_c^2(1+St^2)$), we find not five, but four decay rates that are independent of the wavenumber $k$. Two of them decay again on a timescale equal to the stopping time: 
\begin{equation}\label{eq:large_scale_sol_1}
    \mathrm{Re}[i\omega_{1,2}]=-t_s^{-1}
\end{equation}
Another pair decays on a constant timescale equal to $t_t$:
\begin{equation}\label{eq:large_scale_sol_2}
    \mathrm{Re}[i\omega_{3,4}]=-t_t^{-1}
\end{equation}
For $t_s \gg t_\mathrm{corr}$, the solution in Eq.~\ref{eq:large_scale_sol_1} is indistinguishable from the solution in Eq.~\ref{eq:large_scale_sol_2}. Therefore, the four solutions are represented by only a single sloped line in the upper left corner of the left subplot of Fig.~\ref{fig:2d_disp_relation}.\newline
The fifth solution that we find on large scales is a diffusive solution akin to Eq.~\ref{eq:second_1d_eigenvalue_1} in one dimension and is therefore distinctly different from the other four: 
\begin{equation}\label{eq:small_scale_sol_5}
    i\omega_{5}=-\frac{D k^2}{1+St^2}
\end{equation}
The decay rate in Eq.~\ref{eq:small_scale_sol_5} is proportional to $\propto k^2$ and thus has the properties of a diffusive solution. Notably, compared to the diffusive solution in one dimension, the decay rate is modified by an additional factor $1/(1+St^2)$, i.e., the solution decays more slowly for Stokes numbers above unity, where it scales as $St^{-2}$. This is also the reason the transition from small-scale behavior to large-scale behavior occurs at a wavenumber $k=k_c\sqrt{1+St^2}$, which is a factor of $\sqrt{1+St^2}$ above the characteristic wavenumber $k_c$ (i.e., at smaller scales for $St > 1$).\newline
Based on the result in Eq.~\ref{eq:small_scale_sol_5}, we define the effective radial dust diffusion coefficient as
\begin{equation}\label{eq:radial_dust_diff}
    D_{d,r}^\mathrm{eff} = \frac{D}{1+St^2}
\end{equation}
where $D$ is the diffusion coefficient, calculated using the values of the correlation time $t_\mathrm{corr}$ and the squared turbulent velocity dispersion $u'^2$, as in Eq.~\ref{eq:charact_relation_D_t_corr}. Here, the effective radial dust diffusion coefficient $D_{d,r}^\mathrm{eff}$ is the actual measure of the strength of turbulent dust diffusion in radial direction.  Interestingly, the strength of turbulent diffusion is independent of the correlation time $t_\mathrm{corr}$. As such, this result is in agreement with Eq.~\ref{eq:YL_radial_diff} and confirms the findings of \citet{Youdin2007}.\newline
In Fig.~\ref{fig:eff_diffusivity}, we plot the numerically determined diffusive solution of the dispersion relation of Eq.~\ref{eq:2d_Matrix} as a function of the Stokes number (black line). In Fig.~\ref{fig:eff_diffusivity}, we also plot the exact result of \citet{Youdin2007} regarding radial turbulent transport in an axisymmetric disk (their Eq.~37). For small Stokes numbers ($St\ll 1$), our solution is identical to that of \citet{Youdin2007}. For large Stokes numbers ($St\gg 1$), our solution and the detailed formula of \citet{Youdin2007} have the same scaling ($\propto St^{-2}$) but deviate by a constant factor of an order of unity.

\subsubsection{Physical Interpretation}
\label{eq:Physical Interpretation}
We provide a physical explanation for the reduced strength of radial diffusion in a two-dimensional disk by reiterating the explanation by \citet{Youdin2007}. \newline
We consider the case $t_s \gg t_\mathrm{corr}$ and $St\gg 1$ so that particles decouple from the turbulent motion and also the orbital motion of gas. In a Keplerian disk, these loosely coupled particles undergo epicyclic oscillations with frequency $\Omega$ and length scale $l_\mathrm{epi}=\sqrt{u'^2}/\Omega$. As the particles undergo epicyclic oscillations, they receive short uncorrelated kicks of duration $t_\mathrm{corr}$. An individual particle receives a number of $N=1/(t_\mathrm{corr} \Omega)$ velocity kicks of magnitude $v_\mathrm{kick}\sim\sqrt{u'^2}/(t_s\Omega)$ during an orbital oscillation. Interpreting this as a random walk, the total change in the velocity of a particle during an orbital time is $\delta v\sim v_\mathrm{kick} \sqrt{N}\sim \sqrt{u'^2 t_\mathrm{corr}}/t_s$, which moves the particle a distance of $\delta r\sim \delta v \Omega^{-1}$ every orbital period. A random walk with step size $\delta r$ every orbital period gives a diffusion coefficient $D_{d,r}^\mathrm{eff}\sim \delta r^2 \Omega \sim D/St^2$ as in Eq.~\ref{eq:radial_dust_diff}. For the last equality, we have taken $D=u'^2 t_\mathrm{corr}$ as in Eq.~\ref{eq:charact_relation_D_t_corr}.

\section{Discussion and Summary}
\label{sec:Summary}
In this work, we derive a novel dust turbulent transport model based on a density-weighted mean-field theory and appropriate turbulence closures. The main contribution of this work is a set of mean-field equations that describe the dynamics of dust in protoplanetary disks exhibiting homogeneous and isotropic turbulence. The model is characterized by two parameters, namely the diffusion coefficient $D$ and the correlation time $t_\mathrm{corr}$. \newline
In this paper, we review the popular gradient diffusion approach, for describing the turbulent transport of dust in protoplanetary disks, in Sec. \ref{sec:The Turbulent Particle Diffusion Model}, and highlight the fact that classical gradient diffusion does not guarantee angular momentum conservation in disks. Further, there seems to be no clear consensus on whether the diffused quantity is the absolute dust density $\rho_d$ or the dust concentration relative to gas $\rho_d/\rho_g$. Moreover, orbital effects, that have the potential to reduce the effective strength of diffusion, are not self-consistently captured. Given these limitations, we argue that there is a need for improved transport models that accurately capture the physics of turbulent dust transport in protoplanetary disks. \newline
The model in its most general form describes the averaged dust dynamics with a set of $1+3+3 =7$ partial differential equations (Eqs.~\ref{eq:momentum_like_a}, \ref{eq:momentum_like_b} and \ref{eq:mass_eq_22}). Applying the same averaging approach to the locally isothermal gas equations, we combine them with another set of $1+3=4$ equations (Eqs.~\ref{eq:favre_mass_eq_gas} and \ref{eq:Favre_averaged_mom_gas_2}) to describe the full two-fluid system (gas+dust) in three dimensions with a set of eleven coupled partial differential equations. With Eqs.~\ref{eq:def_of_t_corr} and \ref{eq:charact_relation_D_t_corr}, our mean-field approach provides a method for calibrating the two model parameters for a specific example of turbulence. \newline
Compared to previous models, our model introduces a novel momentum conservation equation that describes the dynamics of the turbulent dust mass flux $\bar{\rho}_d v_i^*$, and is thus capable of capturing non-local turbulent transport effects. In essence, the turbulent dust transport is driven by a turbulent pressure and dissipated by a drag-like term, and as such, the model fully conserves global angular and linear momentum. In the dynamic equilibrium between the driving and dissipating forces, we recover the gradient diffusion model of \citet{Huang22}. In the limit of large particles $t_s\gg t_\mathrm{corr}$ (or equivalently short correlation times) and a uniform gas background ($\rho_g=\mathrm{const.}$, $c_s=\mathrm{const.}$), our dynamical equations are identical to the momentum-conserving model of \citet{Klahr2021}. \newline
We show in Sec.~\ref{eq:Turbulent Particle Transport Beyond Diffusion} that in a balance between the driving and dissipative terms, we recover the classical gradient diffusion solution, with the diffused quantity being the absolute dust density $\rho_d$. However, we further argue that turbulent dust transport should not be considered in isolation. We find that in a nonuniform and static gas background ($\nabla \rho_g \neq 0$ and $\tilde{u}=0$), dust to couple to a mean flow in the gas via the explicit drag term, introducing an additional transport flux to the dust. Consequently, our formalism shows self-consistently that the turbulent dust mass flux in a static gas background ($\tilde{u}=0$) is ultimately governed by the gradient of the dust concentration $\rho_d/\rho_g$. \newline
Furthermore, for large dust particles ($t_s\gg t_\mathrm{corr}$), we find novel turbulent transport flux towards gradients of the stopping time which have not been predicted by previous Eulerian gradient diffusion models. We confirm this by means of a numerical experiment, comparing to the stochastic Lagrangian turbulence model of \citet{Ormel07}.\newline
In the absence of orbital effects and in a steady-state gas background, the total turbulent equilibrium dust flux reads
\begin{equation}\tag{\ref{eq:steady_state_transport_flux}}
    \bar{\rho}_d\tilde{v}_i=-D\rho_g\frac{\partial}{\partial x_i}\frac{\bar{\rho}_d}{\bar{\rho}_g}-D\bar{\rho}_d\frac{\partial}{\partial x_i}\ln t_t^{-1}
\end{equation}
which contains both the gradient diffusion flux and the novel flux contribution. We stress that Eq.~\ref{eq:steady_state_transport_flux} is only valid in equilibrium, i.e., under force balance. \newline
Applying our novel turbulent transport model to study the dust distribution in protoplanetary disks in Sec.~\ref{sec:Vertical Steady-State Disk Profile}, we recover the vertical steady-state profile of \citet{Fromang2009} in the limit of subsonic turbulence $u'^2\ll c_s^2$. Formally, we extend the validity of the solution to large grains (St>1) because in our derivation, we do not invoke the terminal velocity approximation, which in the aforementioned work limited the validity of the solution to small particles ($St\ll 1$). Consequently, we self-consistently reproduce the small particle scaling ($h_d^2/h_g^2 = 1$ for $St\ll 1$) \textit{and} the large particle scaling ($h_d^2 / h_g^2 = \delta / St$ for $St \gg 1$) of the vertical dust scale height without the need for heuristic arguments. \newline
In Sec.~\ref{sec:LPA}, we study the decay of small perturbations to the dust density due to turbulent mixing. We find the turbulent time $t_t=t_s+t_\mathrm{corr}$, to set a lower limit on the decay timescale. For small dust grains ($t_s \ll t_\mathrm{corr}$) this lower limit is equal to the correlation time $t_t \approx t_\mathrm{corr}$, which can, depending on the nature of the underlying turbulence, be comparable to the orbital time. For large grains ($t_s \gg t_\mathrm{corr}$), the lower limit to the decay time is equal to the stopping time $t_s$, which can be larger than the orbital timescales for $St> 1$. \newline
Small-scale perturbations, with a wave number $k$ larger than the characteristic wavenumber $k_c=1/\sqrt{D t_t}$ decay slower (by a factor $k^2/k_c^2$) than a diffusive solution would predict. For values appropriate for protoplanetary disks ($St = 0.025$, $\delta = 10^{-3}$), the characteristic wave number $k_c$, that is, the threshold above which (meaning on larger wavenumbers and smaller physical scales), diffusion is quenched corresponds to a spatial length scale of 20 per cent of the gas scale height $h_g$. At these small scales, perturbations still decay due to drag, but by a factor $Dk^2 t_t$ slower compared to gradient diffusion. \newline
\citet{Umurhan20} showed that gradient diffusion suppresses the smallest modes of the streaming instability. Future work should explore, how the reduction of the strength of diffusion at small scales that we predict affects this result. \newline
In a protoplanetary disk, we find that orbital effects reduce the effective diffusivity of large grains ($St\gtrsim 1$). Specifically, we find the strength of diffusion in both radial and vertical directions to scale as $1/(1+St^2)$ in agreement with the detailed analysis of \citet{Youdin2007} (up to order unity corrections). We emphasize that the effects of orbital dynamics are implicitly captured by our model. We thus expect our model to appropriately capture orbital effects in disk regions where the flow deviates from being purely Keplerian, such as in the vicinity of orbiting planets. \newline
Lastly, our model also offers advantages over classical gradient diffusion models in terms of numerical implementation. Turbulent transport in our model is pressure-driven, allowing the use of standard, locally isothermal fluid solvers to solve the hydrodynamic dust equations. This removes the need to calculate second-order spatial derivatives of the dust density in the gradient diffusion approach, which can be challenging numerically. \newline
In conclusion, we present an improved general Eulerian model of turbulent dust transport in protoplanetary disks. Our model improves upon several limitations of gradient diffusion models, including the conservation of angular momentum, orbital effects, and the functional form of the diffused quantity. By recovering earlier models in special limiting cases, we improve upon the understanding of turbulent dust transport in protoplanetary disks. Future work should extend this model to more complex scenarios like non-homogeneous or anisotropic turbulence and explore their impact on dust transport in protoplanetary disks.

\section*{Acknowledgements}
We thank the anonymous referee for their thorough review and the helpful comments that lead to the improvements of this paper. F.B. thanks T. Birnstiel for fruitful discussions and the diligent review of the manuscript, which significantly improved the quality of this paper. F.B. acknowledges funding from the Deutsche Forschungsgemeinschaft under Ref. no. FOR 2634/1 and under Germany's Excellence Strategy (EXC-2094–390783311).

\section*{Data availability}
The data underlying this article will be shared on reasonable request to the corresponding author.




\bibliographystyle{mnras}
\bibliography{references} 




\appendix
\label{sec:appendix}

\begin{table}
\caption{List of notations.}
\begin{center}
\begin{tabular}{l l l}\hline
  Symbol & Definition/ First use & Description   \\ 
   \hline\hline
 $\Omega$             & $v_\phi / r$ & Orbital angular velocity\\
 $\Omega_K$           & $\sqrt{G M_* / r^3}$ & Keplerian angular velocity\\
 $\Sigma_d$, $\Sigma_g$ & Sec.~\ref{sec:Axisymmetric Keplerian Disks} & Dust, gas surface density\\

 $\alpha$             & Eq.~\ref{eq:alpha_parametrization}    & Shakura-Sunyaev parameter \\
 $\delta$             & Eq.~\ref{eq:turbulent_delta}          & Dimensionless diffusivity\\
 $\delta_{ij}$        & Eq.~\ref{eq:turb_vel_dispersion} & Kronecker delta \\
 $\nu$                  & Sec.~\ref{sec:Characteristics of Disk Turbulence} & Viscosity \\
 $\zeta_t$           & Eq.~\ref{eq:SEOM3}       & Stochastic variable \\
 $\rho_g$, $\rho_d$            & Sec.~\ref{sec:gas_dynamics}, Sec.~\ref{sec:Dust Dynamics}& Dust, gas volume density \\
 $\tau$              & Eq.~\ref{some_equation_with_Int}  & Time-like integration variable\\
 $\tau_e$            & Fig.~\ref{fig:diffusion_sketch}  & E-folding time\\
 $\omega$            &  Sec.~\ref{sec:Characteristics of Disk Turbulence} & Angular frequency \\
 $i\omega$            &  Sec.~\ref{sec:One Dimension Without External Forces} & Growth rate \\
 $-i\omega$            & Sec.~\ref{sec:One Dimension Without External Forces}  & Decay rate \\
\\

 $D$                 & Eq.~\ref{def:diffusion_coefficient}  & Diffusion coefficient \\
 $D_d^\mathrm{eff}$ &    Eq.~\ref{eq:eff_d_d_z}  & Effective dust diffusion coefficient \\
 $\mathfrak{D}(t)$   & Eq.~\ref{eq:def_frak_D}  & Time-dependent diffusion coefficient \\
 $\hat{E}_g(\omega)$ & Eq.~\ref{eq:Energy_spectrum_def}  & Energy spectrum\\
 $J_i$               & Eq.~\ref{eq:trad_adviff_eq_2}  & Turbulent mass flux \\
 $P_{ij}$            & Eq.~\ref{eq:pressure_tensor}  & Turbulent pressure tensor \\
 $R_{ij}$            & Eq.~\ref{eq:Reynolds_tensore}  & Reynolds tensor \\
 $Sc$                & $D_d/D_g$, Eq.~\ref{eq:def_Schmidt_number}  & Schmidt number \\
 $Sc_\mathrm{hydro}$ & $\nu/ D_g$, Sec.~\ref{sec:Characteristics of Disk Turbulence}  & Hydro Schmidt number \\
 $St$                & $t_s\Omega$, Eq.~\ref{eq:Stokes_N_definition}  & Stokes number  \\
 $W_t$               & Eq.~\ref{eq:SEOM3}  & Wiener process \\
\\

 $c_d$               &  Eq.~\ref{eq:turb_vel_disp_def} & Turbulent particle velocity dispersion\\
 $c_s$               &  Eq.~\ref{eq:c_s_T_relation} & Gas sound speed \\
 $g_i$               & Eq.~\ref{eq:gas_mom_cons}  & Gravitational acceleration\\
 $h_g$               &  Eq.~\ref{eq:sch_defintion} & Gas pressure scale height \\
 $h_d$               & Eq.~\ref{eq:ex_sclae_height_ratio} & Dust scale height \\
 $k$                    & Eq.~\ref{eq:pert_rho_lin} & Wave number \\
 $k_c$                    & Eq.~\ref{eq:character_wave_number} & Characteristic wave number \\
 $l_\mathrm{eddy}$   &  Eq.~\ref{eq:eddy_length} & Eddy length\\
 $p$                 & Eq.~\ref{eq_isothermal_eos}  & Thermal pressure \\
 $p_t$               &  Eq.~\ref{eq:iso_turb_pressure} & Isotropic turbulent pressure \\
 $t_s$               &  Eq.~\ref{eq:st_time} & Stopping time \\
 $t_\mathrm{corr}$   &  Eq.~\ref{eq:def_of_t_corr} & Correlation time of turbulence \\
 $t_\mathrm{diff}$  & Eq.~\ref{eq:diffusion timescale} & Diffusion timescale \\
 $t_t$               &  $t_s+t_\mathrm{corr}$, Eq.~\ref{eq_turbulence_time} & Characteristic time of turbulence \\
 $u_i$               & Eq.~\ref{eq:gas_continuity}  & Instantaneous gas velocity\\
 $\delta u$          &  Eq.~\ref{eq:SEOM2} & Stochastic turbulent fluctuation \\
 $v_i$               & Eq.~\ref{eq:lagr_1}  & Instantaneous dust velocity\\
 $\bar{v}_i$         & Eq.~\ref{eq:Reynolds_vel_def}  & Reynolds-averaged dust velocity\\
 $\tilde{v}_i$       &  Eq.~\ref{eq:prop_1} & Favre-averaged dust velocity\\
 $v^*_i$             & Eq.~\ref{eq:diff_velocity}    & Turbulent dust transport velocity\\
 $v'_i$              & Eq.~\ref{eq:Reynolds_vel_def}  & Fluctuation w.r.t. to $\bar{v}$\\
 $v''_i$             &  Eq.~\ref{eq:Favre_average_def} & Fluctuation w.r.t. to $\tilde{v}$\\
 $v_\mathrm{sett}$   & Eq.~\ref{eq:vertical_sett_velocity} &   Vertical settling velocity\\

 \hline
\end{tabular}
\label{table:notations_kap_5}
\end{center}
\end{table}

\bsp	
\label{lastpage}
\end{document}